\documentclass[12pt]{article}
\usepackage{amsmath,amssymb}
\usepackage[dvips]{graphicx}

\begin{document}



\begin{titlepage}

  \begin{normalsize}
  \begin{flushright}
    YITP-01-81\\
    hep-th/0112073\\
    December 2001
  \end{flushright}
  \end{normalsize}
  
  \vspace{1cm}
  
  \begin{Large}
  \begin{center}
  \bf
    Expanded Strings in the Background of NS5-branes \\
    via a M2-brane, a D2-brane and D0-branes
  \end{center}
  \end{Large}
  \vspace{1cm}

\begin{center}
   Yoshifumi Hyakutake
   \footnote{E-mail :hyaku@yukawa.kyoto-u.ac.jp}
   \\
   \vspace{6mm}
   {\it Yukawa Institute for Theoretical Physics, Kyoto University} 
   \\
   {\it Sakyo-ku, Kyoto 606-8502, Japan}
\end{center}

\vspace{2cm}
\begin{center}
  \large{ABSTRACT}
\end{center}

\begin{quote}
\begin{normalsize}
  Classical configurations of a M2-brane, a D2-brane and 
  D0-branes are investigated in the background of an infinite array of
  M5-branes or NS5-branes. 
  On the M2-brane, we discuss three kinds of configurations, such as
  a sphere, a cylinder and a torus-like one. These are stabilized by virtue of 
  the background fluxes of M5-branes.
  The torus-like M2-brane configuration has winding and momentum numbers of 11th 
  direction, and in terms of the type IIA superstring theory,
  this corresponds to a torus-like D2-brane with electric and magnetic fluxes
  on it. We also reproduce the same configuration from a non-abelian Born-Infeld 
  action for D0-branes. It will be a construction of closed strings from D0-branes.
  An electric flux quantization condition on the D2-brane is also discussed
  in terms of D0-branes.
\end{normalsize}
\end{quote}

\end{titlepage}

\setlength{\baselineskip}{18pt}

\section{Introduction}

The type II superstring theories contain $(p\!+\!1)$-form fields called
Ramond-Ramond potentials, and Dirichlet $p$-branes carry charges
corresponding to these fields\cite{Pol}. 
Under the T-duality\cite{KY,SS}, a D$p$-brane which wraps around a compactified circle 
is transformed into a D$(p\!-\!1)$-brane which does not wrap around the circle,
and vice versa\cite{Pol}.
The world-volume theory on D-branes are constructed
in refs.\cite{Lei}-\cite{TR} and such works are extended in a way consistent with
T-duality in ref.\cite{ABB,BR,Mye}.

The strong coupling limit of the type IIA superstring theory is
described by M-theory\cite{HT,Tow,Wit},
and it is pointed out in ref.\cite{Tow2}
that a M2-brane can be constructed as a bound state of infinite number of
D0-branes. Along this line, Banks et al. conjectured that M-theory is equivalent
to matrix quantum mechanics describing a large number $M$ of
D0-branes\cite{BFSS}.
In the BFSS matrix theory, M2-branes are realized 
by giving a non-trivial commutation relation between 
$U(M)$ adjoint scalar fields\cite{Tay}.
The similar fact that a D2-brane is constructed from 
$M$ D0-branes can also be viewed by using a low energy 
effective action of $M$ coincident D0-branes, which is consistent
with the T-duality\cite{Mye}.

In this paper, we study various M2-brane configurations with momentum in the
background of an infinite linear array of M5-branes in $x^{11}$ direction.
Reducing along the $x^{11}$ direction, the background M5-branes are interpreted as
NS5-branes in the type IIA superstring theory, and M2-branes are
identified with D2-branes or strings according to their
configurations. Interestingly, M2 or D2-brane configurations become stable
against collapse by virtue of the background flux. Then we can study
configurations of M2 or D2-brane by employing their world-volume 
theories, and see that both results agree. 
Those configurations are also reconstructed from the
viewpoint of D0-branes. Now let us see the situations in detail.


In refs.\cite{ARS,ARS2,BDS}, configurations of spherical D2-brane are considered 
in the background of $N$ coincident NS5-branes in the type IIA superstring 
theory. The NS5-branes couple magnetically to 
Neveu-Schwarz 2-form field and the flux penetrates $S^3$ in 
the transverse space. Under this background, a D2-brane with $M$ 
units of magnetic flux on it becomes 
stable if it expands into $S^2$ in the $S^3$.
In section \ref{sec:spheD2}, we will reexamine these configuration by
employing the M2-brane action, and
expect that the spherical D2-brane finally becomes a giant graviton
in the background of $AdS_7 \times S^4$. 
The giant graviton considered here
is a spherical M2-brane
moving along a certain $S^1$ in the $S^4$ with $M$ units of momentum\cite{MST}.
And roughly speaking, if the $x^{11}$ direction is identified with the moving one,
the giant graviton would be considered as the spherical D2-brane
with magnetic fluxes on it.
Similar spherical D2-brane configurations in the background of 
D$p$-branes have also been considered in refs.\cite{DJM,Asa}.

In this paper, we also investigate the case where the $x^{11}$ direction is 
identified with one of the world-volume directions of the spherical M2-brane.
After the double dimensional reduction, such M2-brane configurations are
transformed into expanded closed strings moving along a certain
$S^1$ in the $S^3$ in the background of $N$ coincident NS5-branes. 
We will explore this case by using the M2-brane action.
Furthermore, we also examine
the case where such strings are moving along the $x^{11}$ direction
by employing the M2-brane action.
In fact, it is impossible for M2-brane to have momentum along its extending
directions. Therefore in order to assign the momentum of the $x^{11}$
direction, we should slant the world-volume directions of the M2-brane
to the $x^{11}$ direction. In 10-dimensional space-time, the M2-brane
shapes like a torus and,
in terms of type IIA superstring theory, this corresponds 
to a toroidal D2-brane with electric and magnetic flux on it.
As in the case of the BFSS matrix theory or Myers effect,
it is an interesting problem to realize such configurations
from the viewpoint of $M$ D0-branes.
The main purpose of this paper is to exhibit the detailed
relations among M2-brane, D2-brane, fundamental strings and D0-branes.

The outline of this paper is as follows.
In section \ref{sec:spheD2}, the configurations of the spherical D2-brane 
with $M$ units of magnetic flux on it are reviewed from the viewpoint 
of the M2-brane action. The stability of these 
configurations are also discussed.
In section \ref{sec:Expstr}, we explore the configurations of the expanded 
closed string and toroidal D2-brane with electric and magnetic fluxes on it,
by employing the actions of M2-brane or D2-brane. 
In section \ref{sec:D0}, we reconstruct the action of the 
toroidal D2-brane from the Born-Infeld action for $M$ D0-branes.
One important result is that the quantization condition of the electric
flux, corresponding the number of fundamental strings,
is expressed from the degrees of freedom of $M$ D0-branes.
Conclusions and discussions are given in section \ref{sec:Condis}.

\vspace{0.5cm}
\section{Spherical M2-brane in the Background of M5-branes} 
\label{sec:spheD2}

\subsection{NS5-branes background} \label{sec:2-1}

Throughout this paper, various configurations of a M2-brane, a D2-brane and
D0-branes are explored in the background of NS5-branes.
Since M2-branes live in 11-dimensional space-time, the
11-dimensional counterpart of the background of NS5-branes is required.
Thus we briefly review the classical descriptions
of NS5-branes and M5-branes.
We begin with the classical solution of $N$ coincident 
M5-branes in 11-dimensional supergravity theory, and then 
by reducing the 11th direction, derive classical geometry 
of $N$ coincident NS5-branes in type IIA supergravity theory\cite{IMSY,ABKS}.

Bosonic fields of the 11-dimensional supergravity
consist of a graviton $G_{MN}$ and a 3-from field $A_{LMN}$.
Here the capital letters represent 11-dimensional indices.
Then the classical solution of $N$ coincident M5-branes in the 11-dimensional
supergravity is given by a metric of the form
\begin{alignat}{3}
  ds_{11}^2 &= f^{-\frac{1}{3}} \eta_{\mu\nu} dx^\mu dx^\nu
  + f^{\frac{2}{3}} \delta_{ab} dx^a dx^b,
  \quad f = 1 + \frac{\pi N \ell_p^3}{\tilde{r}^3} ,
  \\
  &\quad \tilde{r}^2 = \sum_a (x^a)^2, \quad
  \mu,\nu = 0,1,\cdots,5, \quad a,b=6,7,8,9,11, \notag
\end{alignat}
and a 4-form field strength of the form
\begin{alignat}{3}
  F = dA = 3\pi N\ell_p^3 \, dv_{S^4} . 
\end{alignat}
Here $dv_{S^4}$ denotes the volume form of a unit $S^4$ and 
$\ell_p$ is the Planck length in the 11-dimensional theory.
The $N$ coincident M5-branes are parallel to the $x^{\mu}$
directions and located at $\tilde{r}=0$ in the transverse space.

Our next task is to make a classical solution 
which corresponds to the periodic configuration
of $N$ coincident M5-branes
along the $x^{11}$ direction at intervals of $2\pi R_{11}$. 
This can be done by modifying the harmonic function $f$ as follows:
\begin{alignat}{3}
  &\quad f = 1 + \sum_{n=-\infty}^\infty \frac{\pi N \ell_p^3 }
  {\big(r^2 + (x^{11}-2\pi n R_{11})^2 \big)^\frac{3}{2}} ,
  \quad r^2 = \sum_i (x^i)^2, \quad i = 6,\cdots,9, \label{eq:harmo}
\end{alignat}
and the 4-form field strength in a similar way:
\begin{alignat}{3}
  & F_{ijkl} = 3\pi N \ell_p^3 \epsilon_{ijkl} \sum_{n=-\infty}^\infty
  \frac{x^{11}-2\pi n R_{11}}
  {\big(r^2 + (x^{11}-2\pi n R_{11})^2 \big)^\frac{5}{2}},
  \\
  & F_{ijk\,11} = 3\pi N \ell_p^3 \epsilon_{mijk} \sum_{n=-\infty}^\infty
  \frac{x^m}{\big(r^2 + (x^{11}-2\pi n R_{11})^2 \big)^\frac{5}{2}}. \notag
\end{alignat}
In order to derive the classical description
of $N$ coincident NS5-branes, it is necessary to take a limit of
$1 \ll r/R_{11}$. In this limit, the summation on $n$ is approximated
to an integral and the metric and the 4-form field strength become,
\begin{alignat}{3}
  ds_{11}^2 &= f^{-\frac{1}{3}} \eta_{\mu\nu} dx^\mu dx^\nu
  + f^{\frac{2}{3}} \delta_{ij} dx^i dx^j+ f^{\frac{2}{3}}(dx^{11})^2 ,
  \quad f = 1 + \frac{N \ell_p^3}{R_{11}r^2}, \label{eq:NS5M}
  \\
  F &= \frac{2N\ell_p^3}{R_{11}} dv_{S^3} \wedge dx^{11}. \notag
\end{alignat}
This is an expression for the classical solution of $N$ coincident NS5-branes
in the 11-dimensional supergravity.
In fact, by performing Kaluza-Klein dimensional reduction, the classical
solution (\ref{eq:NS5M}) is transformed into that of the type IIA supergravity:
\begin{alignat}{3}
  ds_{10}^2 &= \eta_{\mu\nu} dx^{\mu} dx^{\nu} + f \delta_{ij} dx^i dx^j, \quad
  f = 1 + \frac{N \ell_s^2}{r^2}, \label{eq:NS5}
  \\
  e^\phi &= f^{1/2} , \quad H = 2N\ell_s^2 dv_{S^3}. \notag
\end{alignat}
Here $\phi$ is a dilaton field and $H$ represents the field strength
of NS-NS 2-form $B$.
The relations such as $R_{11} = \ell_s g_s$ and 
$\ell_p = g_s^{1/3} \ell_s$ were used to derive (\ref{eq:NS5}).

\begin{figure}[htb]
\begin{center}
  \includegraphics[width=12cm,height=8cm,keepaspectratio]{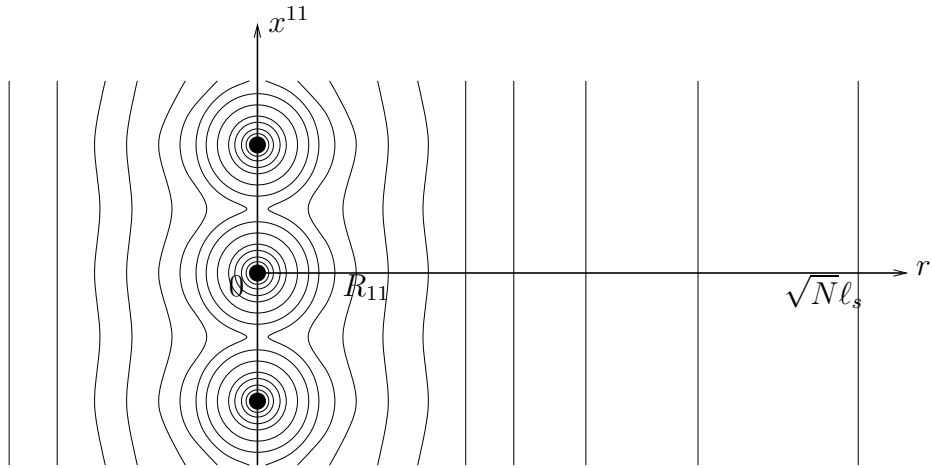}
\begin{picture}(300,0)
  \put(65,80){$0$}
  \put(108,80){$R_{11}$}
  \put(275,77){$\sqrt{N}\ell_s$}
  \put(325,88){$r$}
  \put(80,180){$x^{11}$}
\end{picture}
  \vspace{-0.5cm}
  \caption {A picture of the harmonic function .
  The solid curves represent contour lines of $f$.
  As $r$ decreases to $R_{11}$, asymmetry of the 11th direction 
  will be observed and finally reaches to a throat part of $N$ 
  coincident M5-branes.}
  \label{fig:NS5}
\end{center}
\end{figure}

So far, we have constructed the type IIA supergravity solution for 
$N$ coincident NS5-branes starting with infinite number of M5-branes.
It is clear from the construction that the descriptions (\ref{eq:NS5M}) and
(\ref{eq:NS5}) make sense in the region $1 \ll r/R_{11}$ and $1 \ll N$.
Furthermore, we consider the throat part of the
$N$ coincident NS5-branes. That is the region of 
$1 \ll \sqrt{N}\ell_s/r$ where the constant part of the harmonic function $f$
is negligible. As $r$ approaches to $R_{11}$, the dependence
on the 11th direction will begin to show up (Fig.\ref{fig:NS5}).

In the following sections, the Born-Infeld actions for a D2-brane and D0-branes
are employed, and we must assume $g_s e^{\phi} \ll 1$
in order for the world-volume analyses to be valid.
It is equivalent to the condition
$\sqrt{N} \ll r/R_{11}$, and can be achieved by taking $g_s$
as sufficiently small.

\subsection{Spherical M2-brane in the background of M5-branes}\label{sec:2-2}

The transverse space of $N$ coincident NS5-branes is 4-dimensional
and the flux penetrate $S^3$ in this space.
Since the homotopy group $\pi_2(S^3)$ is trivial, a D2-brane wrapped on $S^2$
in the $S^3$ shrinks to zero size in general.
However, if there are magnetic fluxes on the spherical D2-brane,
it can become stable with some finite radius\cite{BDS}.
In this subsection,
we reexamine configurations of the flux stabilized spherical D2-brane 
in the background of $N$ coincident NS5-branes,
by employing the M2-brane action.

A single M2-brane action is given by the sum of the Nambu-Goto and the
Wess-Zumino type terms of the form
\begin{alignat}{3}
  S_{\text{M2}} &= -T_2 \int d^3 \xi \sqrt{-\det P[G]_{ab}} + T_2 \int P[A],
  \label{eq:M2ac}
\end{alignat}
where $\xi^a$ denote world-volume coordinates of the M2-brane
and $P[\cdots]$ means the 
pullback operation. The tension of the M2-brane is expressed as
$T_2 = 1/4\pi^2 \ell_p^3$, and the
background fields are given by the classical solution (\ref{eq:NS5M})
of M5-branes.
In this subsection, we use the following polar coordinates for
$x^6,\cdots,x^9,x^{11}$ directions:
\begin{alignat}{3}
  &x^6 \!=\! r\sin\theta\sin\theta_1\sin\theta_2 ,\;\,
  x^7 \!=\! r\sin\theta\sin\theta_1\cos\theta_2 ,\;\,
  x^8 \!=\! r\sin\theta\cos\theta_1 ,
  \\
  &x^9 \!=\! r\cos\theta ,\;\, x^{11} \!=\! R_{11}\phi. \notag
\end{alignat}
Parameter regions are given by 
$0 \leq \theta, \theta_1 \leq \pi$ and $0 \leq \theta_2, \phi \leq 2\pi$.
Then the background metric and the 3-form gauge field, 
in adopting an appropriate gauge, are written as follows,
\begin{alignat}{3}
  ds_{11}^2 &= f^{-\frac{1}{3}} \eta_{\mu\nu} dx^\mu dx^\nu
  + f^{\frac{2}{3}} \big(dr^2 + r^2 d\theta^2 + r^2 \sin^2 \theta (d\theta_1^2
  + \sin^2\theta_1 d\theta_2^2) \big) + f^{\frac{2}{3}} R_{11}^2 (d\phi)^2 ,
  \notag
  \\
  A &= N\ell_p^3 (\theta - \sin\theta\cos\theta) \sin\theta_1 
  \; d\theta_1 \wedge d\theta_2 \wedge d\phi,
  \quad f = \frac{N\ell_p^3}{R_{11}r^2}. 
\end{alignat}
What we are interested in here is spherical configurations of the M2-brane,
thus its world-volume coordinates $(\xi^0,\xi^1,\xi^2)$ are chosen as 
$(t,\theta_1,\theta_2)$.
As for the scalar fields which represent the positions of the M2-brane,
we assume that $x^1,\cdots,x^5$ are equal to zero and 
$r$, $\theta$ and $\phi$ are functions of $t$.
Now preparations are complete, we can evaluate
a Lagrangian of the M2-brane straightforwardly.
It becomes
\begin{alignat}{3}
  \mathcal{L}_{\text{M2}} 
  &= -4\pi T_2 f^{\frac{1}{2}} r^2 \sin^2\theta
  \sqrt{1 \!-\! f\dot{r}^2 \!-\! fr^2\dot{\theta}^2 \!-\! 
  fR_{11}^2 \dot{\phi}^2} + \frac{N}{\pi}
  (\theta \!-\! \sin\theta\cos\theta) \dot{\phi} .
\end{alignat}
Note that the momentum conjugate to $\phi$ is a conserved quantity and 
should be quantized since $\phi$ is periodic.
The second term contributes to the potential energy of this system and
decreases it when $\dot{\phi}$ is positive.

In order to find stable configurations of the spherical M2-brane,
it is useful to move to 
the Hamiltonian formalism. As usual, conjugate momenta are defined as
\begin{alignat}{3}
  P_r &\equiv \frac{\partial \mathcal{L}_{\text{M2}}}{\partial \dot{r}}
  = \frac{4\pi T_2 f^{\frac{3}{2}} r^2 \sin^2\theta \,\dot{r}}
  {\sqrt{1 - f\dot{r}^2 - fr^2\dot{\theta}^2 - fR_{11}^2 \dot{\phi}^2}},
  \notag 
  \\
  P_\theta &\equiv \frac{\partial \mathcal{L}_{\text{M2}}}{\partial \dot{\theta}}
  = \frac{4\pi T_2 f^{\frac{3}{2}} r^4 \sin^2\theta \,\dot{\theta}}
  {\sqrt{1 - f\dot{r}^2 - fr^2\dot{\theta}^2 - fR_{11}^2 \dot{\phi}^2}},
  \\
  P_\phi &\equiv \frac{\partial \mathcal{L}_{\text{M2}}}{\partial \dot{\phi}}
  = \frac{4\pi T_2 f^{\frac{3}{2}} r^2 R_{11}^2 \sin^2\theta \, \dot{\phi}}
  {\sqrt{1 - f\dot{r}^2 - fr^2\dot{\theta}^2 - fR_{11}^2 \dot{\phi}^2}}
  + \frac{N}{\pi} \big(\theta - \sin\theta\cos\theta\big),
  \notag 
\end{alignat}
and the Hamiltonian is given by
\begin{alignat}{3}
  \mathcal{H}_{\text{M2}} &\equiv P_r \dot{r} + P_\theta \dot{\theta} 
  + P_\phi \dot{\phi} - \mathcal{L}_{\text{M2}} , \label{eq:M2Ham}
  \\
  &= f^{-\frac{1}{6}} \sqrt{\bigg(\frac{P_r}{f^{\frac{1}{3}}}\bigg)^2 
  + \bigg(\frac{P_\theta}{f^{\frac{1}{3}}r} \bigg)^2
  + \bigg(\frac{P_\phi \!-\! \frac{N}{\pi} 
  (\theta \!-\! \sin\theta\cos\theta)}{f^{\frac{1}{3}}R_{11}}\bigg)^2
  + \big( 4\pi T_2 f^{\frac{2}{3}}r^2 \sin^2\theta \big)^2}. \notag
\end{alignat}
Physical interpretations of this Hamiltonian are as follows.
The coefficient of the square root is the red shift factor.
The first and the second terms in the square root are kinetic energies
of the $r$ and $\theta$ directions,
which are the same as in the case of a point particle.
The third term in the square root corresponds to the momentum along the 11th 
direction reduced by an effect of the background flux of M5-branes. 
The fourth term represents the mass of the spherical M2-brane with the radius
$f^{\frac{1}{3}}r\sin\theta$.
If there is no background flux, the energy of the M2-brane reaches 
minima when $\theta$ is equal to $0$ or $\pi$, which represent contracting
M2-branes. Thus an essence of flux stabilization comes from 
the existence of the third term.

Now let us solve the equations of motion
obtained from the Hamiltonian (\ref{eq:M2Ham}).
The equations of motion become of the forms
\begin{alignat}{3}
  \dot{r} &= \frac{P_r}{f\mathcal{H}_{\text{M2}}}, \qquad 
  \dot{P}_r = -\frac{1}{fr\mathcal{H}_{\text{M2}}} \bigg[ P_r^2
  + \bigg(\frac{P_\phi - \frac{N}{\pi}(\theta - \sin\theta\cos\theta)}
  {R_{11}} \bigg)^2 + \frac{N^2}{\pi^2 R_{11}^2} \sin^4\theta \bigg], \notag
  \\
  \dot{\theta} &= \frac{P_\theta}{fr^2\mathcal{H}_{\text{M2}}},\quad 
  \dot{P}_\theta = \frac{2N\sin^2\theta}
  {\pi f R_{11}^2\mathcal{H}_{\text{M2}}}
  \Big(P_\phi - \frac{N}{\pi} \theta \Big),
  \\
  \dot{\phi} &= \frac{1}{fR_{11}^2\mathcal{H}_{\text{M2}}}
  \Big( P_\phi - \frac{N}{\pi}(\theta - \sin\theta\cos\theta) \Big), \quad 
  \dot{P}_\phi = 0. \notag
\end{alignat}
As already mentioned, $P_\phi$ takes an integer $M$.
In order to solve these equations, we make an assumption 
that $\dot{\theta}=0$. This requires $P_\theta=0$ and
hence $\dot{P_\theta}=0$.
Then there are three kinds of solutions. Two of those are
\begin{alignat}{3}
  \theta = 0, \quad \mathcal{H}_{\text{M2}} = 
  \sqrt{ \frac{P_r^2}{f} + \frac{M^2}{f R_{11}^2}}, \label{eq:north}
\end{alignat}
and
\begin{alignat}{3}
  \theta = \pi, \quad \mathcal{H}_{\text{M2}} = 
  \sqrt{ \frac{P_r^2}{f} + \frac{(M-N)^2}{f R_{11}^2}}. \label{eq:south}
\end{alignat}
The former represents a shrinking M2-brane at the north pole of 
the $S^3$ and likewise the latter at the south pole.
The remaining solution is given by
\begin{alignat}{3}
  &\theta = \frac{M}{N}\pi, \quad \mathcal{H}_{\text{M2}} =
  \sqrt{ \frac{P_r^2}{f}
  + \frac{N^2\sin^2(\frac{M}{N}\pi)}{\pi^2 f R_{11}^2} }. 
  \label{eq:sphe}
\end{alignat}
Note that this solution exists only in the region of $0 \le M \le N$.
This solution represents a spherically expanded
M2-brane with the radius $f^{\frac{1}{3}}r\sin(\frac{M}{N}\pi)$ and
gives the lowest energy among the three solutions.

If $P_r$ is equal to zero, 
the energy of (\ref{eq:sphe}) can be rewritten as
\begin{alignat}{3}
  \mathcal{H}_{\text{M2}} &= \frac{N}{\pi \ell_s g_s e^\phi}\sin 
  \Big( \frac{M}{N}\pi \Big), \label{eq:spene}
\end{alignat}
where the equation (\ref{eq:NS5}) is used. 
This gives the same energy as a spherical D2-brane of ref.\cite{BDS}. 
In any solution we should also solve the equations
for the radial direction of the forms
\begin{alignat}{3}
  \dot{r} = \frac{P_r}{f\mathcal{H}_{\text{M2}}}, \qquad
  \quad \dot{P}_r = -\frac{\mathcal{H}_{\text{M2}}}{r}. \label{eq:rdire}
\end{alignat}
We will analyze this equation in the next subsection.
\begin{figure}[tb]
  \begin{center}
  \includegraphics[width=4cm,height=4cm,keepaspectratio]{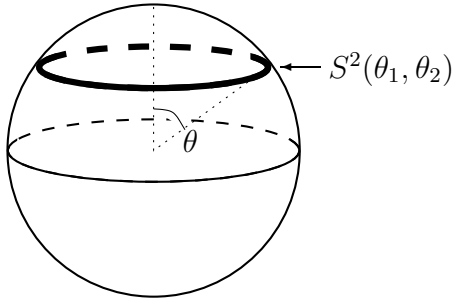}
  \caption{A picture of $S^3$. A spherical M2-brane expands into 
  the bold `sphere'.}
  \begin{picture}(200,0)
    \put(113,95){$\theta$}
    \put(165,127){\vector(-1,0){15}}
    \put(168,123){$S^2 (\theta_1, \theta_2)$}
  \end{picture}
  \label{fig:sphere}
  \end{center}
\end{figure}

\subsection{Stability on the spherical M2-brane}

In ref.\cite{BDS}, the solution (\ref{eq:sphe}) has been shown to be stable
against fluctuations for $\theta$ direction.
To check the full stability of the spherical M2-brane, however, 
we should also take care of the motion in the $r$ direction.

Now let us solve the equations of motion (\ref{eq:rdire}). The Hamiltonian 
(\ref{eq:M2Ham}) is a conserved quantity since
it does not depend on $t$ explicitly, so
the equations of motion can be rewritten as
\begin{alignat}{3}
  \ddot{r} - \frac{2}{r} (\dot{r})^2 + \frac{r}{N\ell_s^2} = 0.
\end{alignat}
Then a solution of this equation is expressed as
\begin{alignat}{3}
  r &= \frac{r_0}{\cosh\big(\frac{t}{\sqrt{N}\ell_s}\big)}. \label{eq:solr}
\end{alignat}
The integral constants are chosen so as to satisfy 
$r(0)=r_0$ and $\dot{r}(0)=0$.
Thus as time evolves the spherical M2-brane falls deep into the
horizon of the infinite linear array of M5-branes.
$P_r$, $\mathcal{H}_{\text{M2}}$ and $\dot{\phi}$ are given by
\begin{alignat}{3}
  P_r &= - \frac{N \sin\theta}{\pi R_{11}} \sinh 
  \Big( \frac{t}{\sqrt{N}\ell_s} \Big), \quad
  \mathcal{H}_{\text{M2}} = \frac{r_0 \sqrt{N}\sin\theta}{\pi\ell_s R_{11}}, \notag
  \\
  \dot{\phi} &= \frac{r_0 \cos\theta}{\sqrt{N}\ell_s R_{11}}
  \frac{1}{\cosh^2 \big( \frac{t}{\sqrt{N}\ell_s} \big)}. \label{eq:solP}
\end{alignat}

As already mentioned in the subsection \ref{sec:2-1}, 
our calculation is reliable in the region of 
$R_{11}/r \ll 1 \ll \sqrt{N}\ell_s/r$ (and $g_s \ll 1$), 
therefore dynamics out of this parameter region is beyond our scope.
One can guess, however, that as $r$ approaches to $R_{11}$
the dependence of the background on the 11th direction is observed, 
and the spherical M2-brane approaches
the throat part of the $N$ coincident M5-branes. 
There the spherical M2-brane could be identified with the giant 
graviton in the background of $AdS_7 \times S^4$.

In the remainder of this subsection we discuss whether the spherical
M2-brane is supersymmetric or not. 
A strategy employed here is the same as that of refs.\cite{GMT,HHI}.
First we search for killing spinors in the background of
the infinite linear array of M5-branes, and next count the number of the killing
spinors which can be zero by $\kappa$-symmetry on the M2-brane.

In the 11-dimensional supergravity theory, 
a variation of a gravitino $\Psi_M$
under supersymmetric transformation is given by the form
\begin{alignat}{3}
  \delta \Psi_M &= \mathcal{\hat{D}}_M \epsilon - \frac{1}{288}
  ({\Gamma_M}^{PQRS} - 8 \delta_M^P \Gamma^{QRS})F_{PQRS} \, \epsilon ,
  \label{eq:killsp}
\end{alignat}
where capital letters denote 11-dimensional space-time 
indices and $\epsilon$ is a Majorana spinor. 
$\mathcal{\hat{D}}_M$ denotes a supercovariant derivative for Majorana spinors.
The full expression for it contains terms of higher order in $\Psi_M$,
but it is just given by an ordinary covariant derivative if we insert
the classical solution (\ref{eq:NS5M}).
$\Gamma_M$ are gamma matrices with space-time indices and
$\Gamma_{\hat{M}}$ are those with local Lorentz indices.
If we substitute the background fields (\ref{eq:NS5M}) for the equation
(\ref{eq:killsp}), the killing spinor equations 
for $M=0,1,\cdots,5,r$ become
\begin{alignat}{3}
  \mathcal{D}_M \epsilon - \frac{f^{\frac{1}{6}}}
  {6\sqrt{N}\ell_s} \Gamma_M \gamma \,\epsilon = 0,
\end{alignat}
and those for $M=\theta,\theta_1,\theta_2,11$ are written as
\begin{alignat}{3}
  \mathcal{D}_M \epsilon + \frac{f^{\frac{1}{6}}}
  {3\sqrt{N}\ell_s} \Gamma_M \gamma \,\epsilon = 0,
\end{alignat}
where $\gamma \equiv \Gamma^{\hat{\theta}\hat{\theta_1}\hat{\theta_2}\hat{11}}$.
The solution for these killing spinor equations is given by
\begin{alignat}{3}
  \epsilon &= r^{\frac{1}{6}} e^{-\frac{1}{2}\theta \Gamma^{\hat{\theta}}
  \gamma}
  e^{-\frac{1}{2} \theta_1 \Gamma^{\hat{\theta_1}\hat{\theta}}} 
  e^{-\frac{1}{2} \theta_2 \Gamma^{\hat{\theta_2}\hat{\theta_1}}} 
  \frac{1+\Gamma^{\hat{r}}\gamma}{2} \epsilon_0,
\end{alignat}
where $\epsilon_0$ is an arbitrary constant Majorana spinor.
The non-trivial part which is expressed by the parameters 
$\theta$, $\theta_1$ and $\theta_2$ corresponds to killing spinors
of $S^3$\cite{LPR}. The background of $N$ coincident NS5-branes
preserves half of space-time supersymmetry, as expected.

Our next task is to count the number of the killing spinors 
which is consistent with the condition of unbroken supersymmety 
on the M2-brane.
A supersymmetric variation of Majorana spinor $\Theta$
on the M2-brane with the $\kappa$-symmetry is given by the form
\begin{alignat}{3}
  \delta \Theta = \epsilon|_{\text{M2}} + (1+\Gamma) \kappa. \label{eq:kap}
\end{alignat}
Here $\Gamma$ is defined, by setting $\Theta=0$, as
\begin{alignat}{3}
  \Gamma \equiv \frac{1}{3! \sqrt{-\det P[G]_{ab}}}
  \epsilon^{abc} \partial_a X^L \partial_b X^M \partial_c X^N \Gamma_{LMN}.
\end{alignat}
This satisfies the property of projection matrix, that is $\Gamma^2=1$.
By using this, the condition $\delta \Theta = 0$ can be written in the form
\begin{alignat}{3}
  (1 - \Gamma) \epsilon|_{\text{M2}} = 0. \label{eq:kap2}
\end{alignat}
Then by substituting the solutions (\ref{eq:solr}) and (\ref{eq:solP})
for $\Gamma$, an equation of the form
\begin{alignat}{3}
  1-\Gamma &= -\frac{1}{\sin\theta} \Gamma^{\hat{\theta}}\gamma
  \Big[ e^{\theta \Gamma^{\hat{\theta}}\gamma} + 
  \cosh\big(\tfrac{t}{\sqrt{N}\ell_s} \big) \Gamma^{\hat{t}\hat{11}}
  + \sinh\big(\tfrac{t}{\sqrt{N}\ell_s} \big) \Gamma^{\hat{r}\hat{11}}
  \Big],
\end{alignat}
is obtained, and the condition (\ref{eq:kap2}) reduces to the form
\begin{alignat}{3}
  \frac{1+ \cosh\big(\frac{t}{\sqrt{N}\ell_s} \big) \, \Gamma^{\hat{t}\hat{11}}
  + \sinh\big(\frac{t}{\sqrt{N}\ell_s} \big) \, \Gamma^{\hat{r}\hat{11}}}{2}
  \frac{1+\Gamma^{\hat{r}}\gamma}{2} \epsilon_0 = 0.
\end{alignat}
Let us operate $\Gamma^{\hat{r}}\gamma$ from the left side of this equation.
Then a similar equation, but with a minus sign in front of the $\cosh$, is obtained.
By subtracting the latter from the former, we finally obtain
a condition of the form
\begin{alignat}{3}
  \frac{1+\Gamma^{\hat{r}}\gamma}{2} \epsilon_0 = 0.
\end{alignat}
From this it follows that the configurations of the spherical M2-brane break the
11-dimensional space-time supersymmetry completely.
This is in contrast with the spherical M2-brane 
in the background of $AdS_7 \times S^4$, where the background has
32 maximal supersymmetry and the spherical M2-brane preserves
half of them\cite{GMT,HHI}. In the frame work of the Wess-Zumino-Witten
model, configurations of spherical D2-brane which preserve 
half of the target space supersymmetry are also discussed\cite{HNS}.

\section{Cylindrical and Torus-like M2-branes in the Background of 
M5-branes} \label{sec:Expstr}

\subsection{Cylindrical M2-brane as closed strings}\label{sec:3-1}

In the previous section, we studied the properties of a spherical M2-brane
in the background of the infinite linear array of M5-branes. 
The 11-dimensional counterpart of this object
is the giant graviton
in the background of $AdS_7 \times S^4$.
Let us see this correspondence in more detail.
Let $N$ be the number of units of 4-form flux on the $S^4$, and the unit $S^4$ be
parametrized by polar coordinates $(\theta,\phi,\theta_1,\theta_2)$ as
\begin{alignat}{3}
  {ds_{S^4}}^2 &= d\theta^2 + \cos^2\theta d\phi^2 + \sin^2\theta
  (d\theta_1^2 + \sin^2\theta_1 d\theta_2^2),
\end{alignat}
where $0 \leq \theta \leq \tfrac{\pi}{2}$, 
$0 \leq \theta_1 \leq \pi$ and $0 \leq \phi, \theta_2 \leq 2\pi$.
M2-brane is wrapped on the $S^2$ parametrized by $(\theta_1,\theta_2)$
and moving in the $\phi$ direction. When a momentum conjugate to $\phi$
takes an integer value $M (\leq N)$, the spherical M2-brane becomes 
stable if $\sin\theta$ is equal to $0$ or $\frac{M}{N}$.
The former solution represents an ordinary graviton and the latter one is
called the giant graviton.
If we identify the $\phi$ direction with the $x^{11}$ direction and do
the dimensional reduction, the giant graviton is expected to become 
a spherical D2-brane with $M$ units of magnetic flux on it,
which was just argued in the previous section.

\begin{figure}[bth]
\begin{center}
  \includegraphics[width=14cm,height=5cm,keepaspectratio]{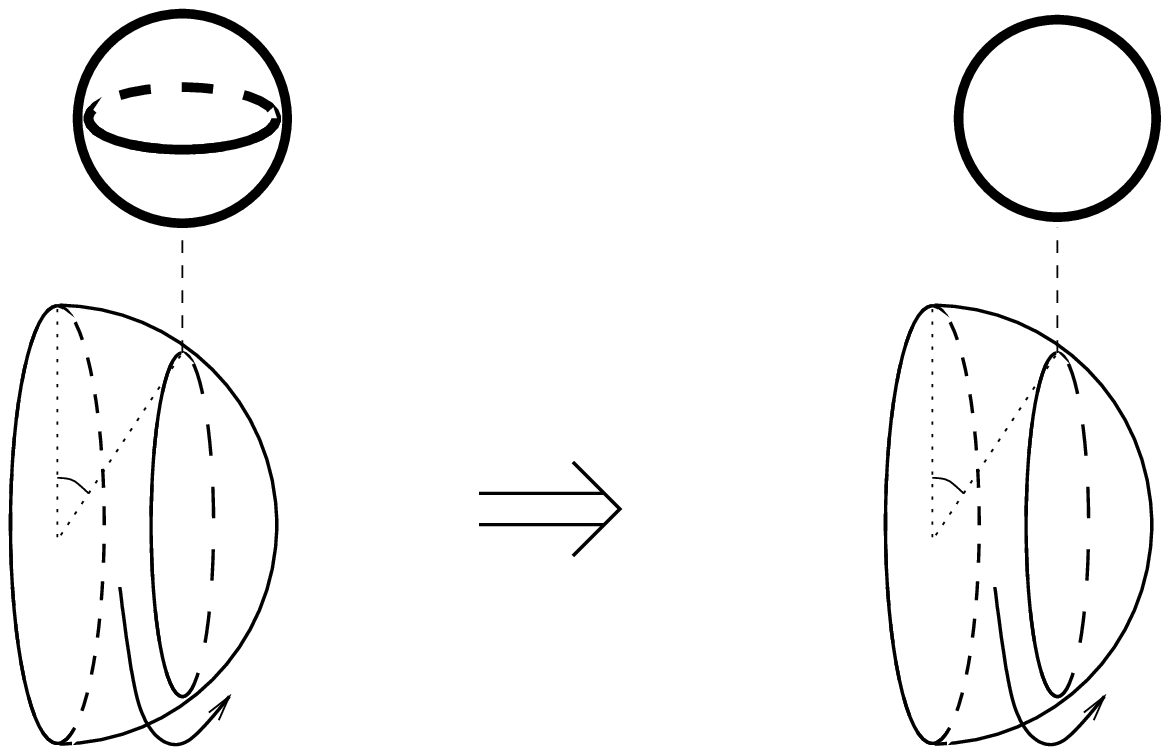}
\begin{picture}(300,0)
  \put(80,-5){$\mathbf{S^4}$}
  \put(67,60){$\theta$}
  \put(100,13){$\phi$}
  \put(60,140){$S^2$}
  \put(40,128){$(\theta_1,\theta_2)$}
  \put(245,-5){$\mathbf{S^3}$}
  \put(230,60){$\theta$}
  \put(263,13){$\phi$}
  \put(226,140){$S^1$}
  \put(222,128){$(\chi)$}
  \put(145,40){$x^{11}\sim \theta_2$}
\end{picture}
  \vspace{0cm}
  \caption{Left hand side represents the giant graviton, and 
  right hand side does an expanded closed string, both moving 
  in the $\phi$ direction.}
  \label{fig:string}
\end{center}
\end{figure}

On the other hand, we can also examine a situation where the 
$x^{11}$ direction is identified with the $\theta_2$ direction,
which is one of the world-volume directions of the giant graviton.
Then the giant graviton will become a closed string, or
a cylindrical M2-brane, in the background
of $N$ coincident NS5-branes (Fig.\ref{fig:string}). 
We will investigate these configurations of an M2-brane in this subsection.

A starting point is the M2-brane action (\ref{eq:M2ac})
in the background of the infinite linear array of M5-branes (\ref{eq:NS5M}).
We adopt the polar coordinates on $x^6,\cdots,x^9$ as in the
Fig.(\ref{fig:string}):
\begin{alignat}{3}
  x^6 \!=\! r\cos\theta\cos\phi ,\quad
  x^7 \!=\! r\cos\theta\sin\phi ,\quad
  x^8 \!=\! r\sin\theta\cos\chi , \quad
  x^9 \!=\! r\sin\theta\sin\chi , \label{eq:S3}
\end{alignat}
where $0 \leq \theta \leq \frac{\pi}{2}$ and 
$0 \leq \phi, \chi \leq 2\pi$.
Then the background metric and the 3-form gauge field,
in an appropriate gauge, become
\begin{alignat}{3}
  ds_{11}^2 &= f^{-\frac{1}{3}} \eta_{\mu\nu} dx^\mu dx^\nu
  + f^{\frac{2}{3}} \big(dr^2 + r^2 d\theta^2 + r^2 \cos^2 \theta d\phi^2
  + r^2 \sin^2\theta d\chi^2 \big) + f^{\frac{2}{3}}(dx^{11})^2 , \notag
  \\
  A &= \frac{N\ell_p^3}{R_{11}} \sin^2\theta d\phi \wedge d\chi \wedge dx^{11},
  \quad f = \frac{N\ell_p^3}{R_{11} r^2}. \label{eq:bac}
\end{alignat}
What we are interested in here is the cylindrical M2-brane extending 
to the $x^{11}$ direction, so we identify the world-volume coordinates
$(\xi^0,\xi^1,\xi^2)$ with the space-time coordinates $(t,\chi,\psi)$.
$\psi$ has periodicity of $2\pi$, and is related to $x^{11}$ as
$x^{11}=L R_{11} \psi$. $L$ is the winding number.
As for the scalar fields which correspond to the positions of the M2-brane,
we assume that $x^1,\cdots,x^5$ are zero and $r$, $\theta$ and $\phi$ 
are functions only of $t$. Then, the metric induced on the M2-brane 
and the pullback of the 3-form gauge field become
\begin{alignat}{3}
  P[G]_{ab} &= 
  \begin{pmatrix}
    -f^{-\frac{1}{3}}(1-f\dot{r}^2-fr^2\dot{\theta}^2-fr^2\cos^2\theta
    \dot{\phi}^2) & 0 & 0 \\
    0 & f^{\frac{2}{3}} r^2 \sin^2\theta & 0 \\
    0 & 0 & f^{\frac{2}{3}}L^2 R_{11}^2
  \end{pmatrix}, \notag
  \\[0.2cm]
  P[A] &= L N \ell_p^3 \sin^2\theta \dot{\phi} \;
  dt \wedge d\chi \wedge d\psi.
\end{alignat}
By substituting the above data into the action (\ref{eq:M2ac}), the Lagrangian
\begin{alignat}{3}
  \mathcal{L}_{\text{M2}} &= - 4\pi^2 T_2 f^{\frac{1}{2}} r L R_{11} \sin\theta
  \sqrt{1-f\dot{r}^2-fr^2\dot{\theta}^2-fr^2\cos^2\theta\dot{\phi}^2}
  + L N \sin^2\theta \dot{\phi}, \label{eq:stlag}
\end{alignat}
is obtained.
It is understood that a momentum conjugate to $\phi$ is a discrete
conserved quantity, and the second term plays the role of the potential
energy which is negative if $\dot{\phi}$ is positive.

So as to examine the potential energy of the configurations
of cylindrical M2-brane, it is appropriate to move
to the Hamiltonian formalism. 
The canonical conjugate momenta are defined as
\begin{alignat}{3}
  P_r &\equiv \frac{\partial \mathcal{L}_{\text{M2}}}{\partial \dot{r}} =
  \frac{4\pi^2 T_2 f^{\frac{3}{2}} r L R_{11} \sin\theta \, \dot{r}}
  {\sqrt{1-f\dot{r}^2-fr^2\dot{\theta}^2-fr^2\cos^2\theta\dot{\phi}^2}}, \notag
  \\
  P_\theta &\equiv \frac{\partial \mathcal{L}_{\text{M2}}}{\partial \dot{\theta}} =
  \frac{4\pi^2 T_2 f^{\frac{3}{2}} r^3 L R_{11} \sin\theta \, \dot{\theta}} 
  {\sqrt{1-f\dot{r}^2-fr^2\dot{\theta}^2-fr^2\cos^2\theta\dot{\phi}^2}},
  \\
  P_\phi &\equiv \frac{\partial \mathcal{L}_{\text{M2}}}{\partial \dot{\phi}} =
  \frac{4\pi^2 T_2 f^{\frac{3}{2}} r^3 L R_{11} \sin\theta \cos^2\theta 
  \, \dot{\phi}}
  {\sqrt{1-f\dot{r}^2-fr^2\dot{\theta}^2-fr^2\cos^2\theta\dot{\phi}^2}}
  + L N \sin^2\theta, \notag
\end{alignat}
and the Hamiltonian is given by the form
\begin{alignat}{3}
  \mathcal{H}_{\text{M2}} &= f^{-\frac{1}{6}}
  \sqrt{\bigg( \frac{P_r}{f^{\frac{1}{3}}} \bigg)^2 + 
  \bigg(\frac{P_\theta}{f^{\frac{1}{3}}r} \bigg)^2 + 
  \bigg( \frac{P_\phi-LN\sin^2\theta}{f^{\frac{1}{3}}r\cos\theta} \bigg)^2 + 
  \big( 4\pi^2 T_2 f^{\frac{2}{3}} r L R_{11} \sin\theta \big)^2}.
  \label{eq:Ham1}
\end{alignat}
Physical interpretations of this Hamiltonian are as follows.
The coefficient of the square root is the red shift factor.
The first and second terms in the square root are the kinetic energies
of $r$ and $\theta$ directions, like those of a point particle.
The third term corresponds to the momentum conjugate to the 
$\phi$ direction shifted by the effect of the background flux of M5-branes. 
The fourth term represents the mass of the cylindrical M2-brane with 
the radius of $f^{\frac{1}{3}}r\sin\theta$.
The main difference from the case of the spherical M2-brane is
that, by multiplying the red shift factor, the fourth term 
does not depend on $r$.
This is due to the fact that the cylindrical M2-brane is just a closed
string and its tension does not depend on the dilaton field $\phi$.

Our next task is to solve the equations of motion obtained from the
Hamiltonian (\ref{eq:Ham1}). They are given by
\begin{alignat}{3}
  &\dot{r} = \frac{P_r}{f \mathcal{H}_{\text{M2}}}, \qquad
  &&\dot{P_r} = -\frac{P_r^2}{fr \mathcal{H}_{\text{M2}}}, \notag
  \\
  &\dot{\theta} = \frac{P_\theta}{fr^2 \mathcal{H}_{\text{M2}}}, \qquad
  &&\dot{P_\theta} = - \frac{\sin\theta}{fr^2 \cos^3\theta \mathcal{H}_{\text{M2}}}
  (P_\phi-LN)^2 ,
  \\
  &\dot{\phi} = \frac{P_\phi-LN\sin^2\theta}
  {fr^2 \cos^2\theta \, \mathcal{H}_{\text{M2}}},
  \qquad\qquad &&\dot{P_\phi} = 0. \notag
\end{alignat}
$P_\phi$ takes an integer value.
Differently from the case of the spherical M2-brane, we can assume that
$\dot{r} = \dot{\theta} = 0$.
These require $P_r=P_\theta=0$ and hence $\dot{P_r}=\dot{P}_\theta=0$.
Then there are two kinds of solutions.
The first one is given by
\begin{alignat}{3}
  \theta = 0, \quad \mathcal{H}_{\text{M2}} = \frac{P_\phi}{f^{\frac{1}{2}}r},
\end{alignat}
where $P_\phi$ is an arbitrary integer.
This is a singular solution and represents a M2-brane shrinking 
like a thin wire. The another solution is written as
\begin{alignat}{3}
  P_\phi = LN, \quad \mathcal{H}_{\text{M2}} = \frac{LN}{f^{\frac{1}{2}}r},
\end{alignat}
where $\theta$ takes an arbitrary value in the region 
$0 \leq \theta \leq \frac{\pi}{2}$.
This solution represents a cylindrical M2-brane with the radius
$f^{\frac{1}{3}}r \sin\theta$.
In terms of type IIA superstring theory, 
this corresponds to $L$ coincident closed strings 
that look like $S^1$ embedded in $S^3$ and have the radius
$\sqrt{N}\ell_s \sin\theta$.
Note that when $\theta=\frac{\pi}{2}$
the $\phi$ direction is not well-defined, so the solution becomes singular.

\begin{figure}[tb]
  \begin{center}
    \includegraphics[width=14cm,height=8cm,keepaspectratio]{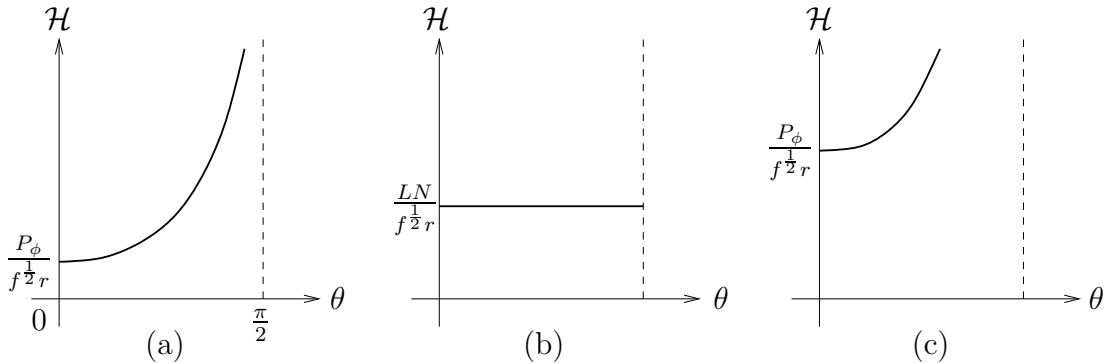}
  \begin{picture}(400,10)
    \put(7,128){$\mathcal{H}$}
    \put(115,22){$\theta$}
    \put(2,14){$0$}
    \put(85,14){$\frac{\pi}{2}$}
    \put(-8,38){$\frac{P_\phi}{f^{\frac{1}{2}}r}$}
    \put(45,5){(a)}
    \put(150,128){$\mathcal{H}$}
    \put(260,22){$\theta$}
    \put(138,59){$\frac{LN}{f^{\frac{1}{2}}r}$}
    \put(188,5){(b)}
    \put(295,128){$\mathcal{H}$}
    \put(402,22){$\theta$}
    \put(280,80){$\frac{P_\phi}{f^{\frac{1}{2}}r}$}
    \put(335,5){(c)}
  \end{picture}
  \caption{The three cases of $0 \leq P_\phi < LN$, $P_\phi = LN$ and
  $LN < P_\phi$ are drawn in (a), (b) and (c) respectively.}
  \label{fig:fig1}
  \end{center}
\end{figure}

Let us investigate these solutions in more detail.
By setting $P_r = P_\theta = 0$, the Hamiltonian (\ref{eq:Ham1}) takes
the form
\begin{alignat}{3}
  \mathcal{H}_{\text{M2}} &= \frac{1}{f^{\frac{1}{2}}r}
  \sqrt{P_\phi^2 + \tan^2\theta (P_\phi-LN)^2}.\label{eq:strpot}
\end{alignat}
The Hamiltonian is drawn as a function of $\theta$ in Fig.\ref{fig:fig1}.
For any value of $P_\phi$, the singular configurations at $\theta = 0$ take
minimal energy, which are just the same energy as that of a point particle moving 
along the equatorial direction in the $S^3$.
The special case occurs when $P_\phi = LN$.
There the cylindrical M2-brane freely expands into an arbitrary 
radius while keeping the minimal energy\footnote{If we take into
account the constant part of the harmonic function $f$, an extra term
of the order $\frac{r^2}{N\ell_s^2}$ might appear in the square root
of the eq.(\ref{eq:strpot}). Then fig.\ref{fig:fig1} is slightly
modified and in the case of $P_{\phi}=LN$, the Hamiltonian becomes a
monotonously increasing function of $\theta$.
There is also a possibility that the back-reaction modifies the
behavior of the Hamiltonian\cite{HHI,Ben}.}

We can reproduce the Lagrangian (\ref{eq:stlag})
by beginning with the Nambu-Goto action for strings of the form
\begin{alignat}{3}
  S &= - LT \int d^2\sigma \sqrt{-\det P[G]_{ab}} + LT \int P[B],
\end{alignat}
where the background is given by the equation (\ref{eq:NS5}).
We also obtain similar kinds of expanded D1-branes 
in the background of $N$ coincident D5-branes.

\subsection{Torus-like M2-brane with winding and momentum numbers
for the 11th direction} \label{sec:3-2}

In the subsection \ref{sec:2-2},
we argued the configurations of the spherical M2-brane with $M$ units of
momentum in the 11th direction.
And in the subsection \ref{sec:3-1},
we discussed the configurations of the cylindrical
M2-brane winding $L$ times around the 11th direction.
In terms of type IIA superstring theory,
the former corresponds to a spherical D2-brane with $M$ 
units of magnetic flux
on it and the latter to $L$ coincident closed strings
moving along the transverse direction,
both in the background of $N$ coincident NS5-branes.

\begin{figure}[bt]
\begin{center}
  \includegraphics[width=14cm,height=5cm,keepaspectratio]{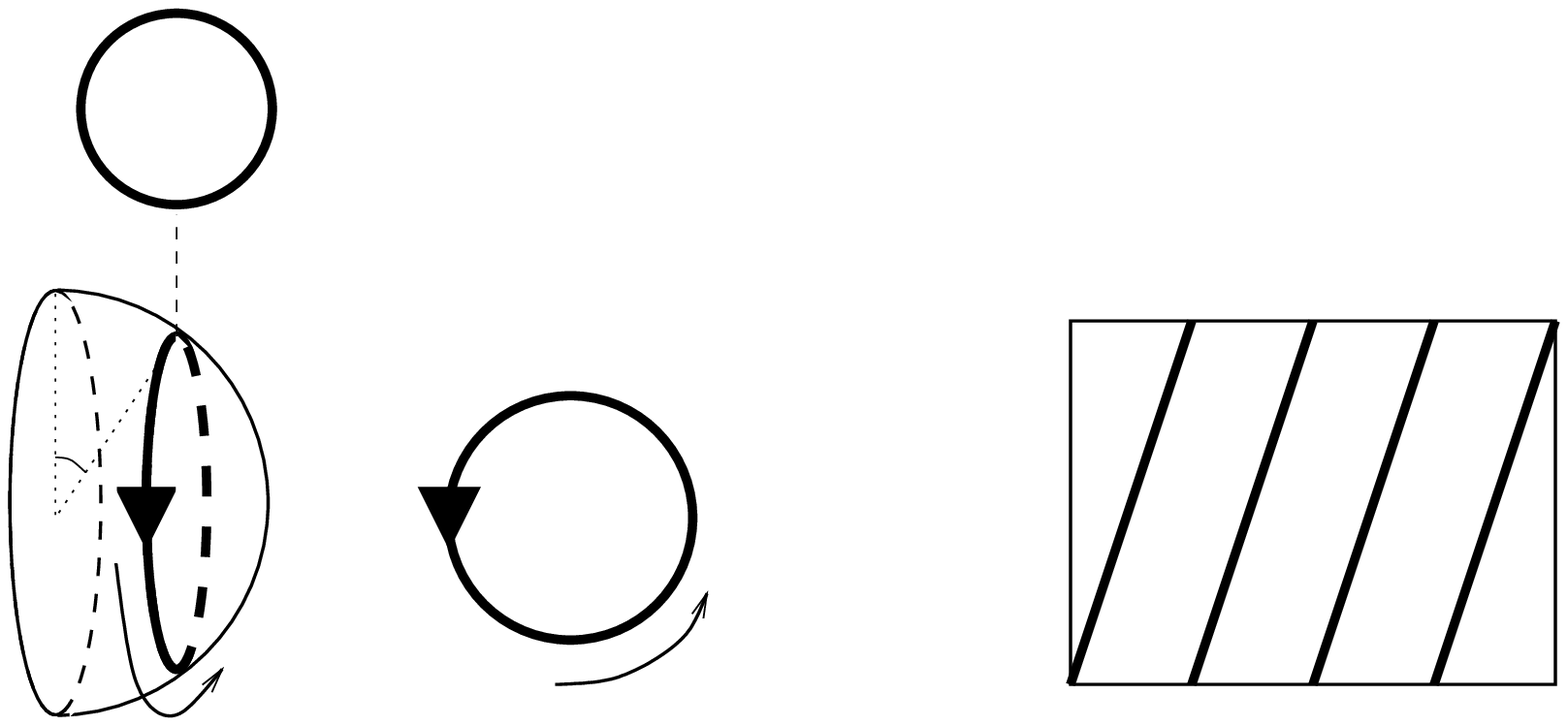}
\begin{picture}(400,0)
  \put(92,-5){$\mathbf{S^3}$}
  \put(137,-5){$\mathbf{\times}$}
  \put(73,60){$\theta$}
  \put(107,13){$\phi$}
  \put(177,-5){$\mathbf{S^1}$}
  \put(207,33){$x^{11}$}
  \put(69,140){$S^1$}
  \put(65,128){$(\chi)$}
  \put(320,13){$\phi$}
  \put(260,57){$x^{11}$}
\end{picture}
  \vspace{0cm}
  \caption{A picture of $S^3 \times S^1$. The torus-like M2-brane
  is wrapped around the $\phi$ and $\chi$ directions.
  M2-brane world-volume wraps onto the $x^{11}$ direction
  $L$ times as it goes around the $\phi$ direction once.}
  \label{fig:torus}
\end{center}
\end{figure}

In this subsection, we investigate configurations of a M2-brane 
with winding and momentum numbers along the 11th direction,
in the background of the infinite array of M5-branes.
In order to have the winding number, the M2-brane should shape like a
torus (Fig.\ref{fig:torus}).

We adopt the polar coordinates on $x^6,\cdots,x^9$ 
like the equations (\ref{eq:S3}) 
in the previous subsection.
The background metric and the 3-form gauge field, in
an appropriate gauge, are given by the equations (\ref{eq:bac}).
Now we want to realize the torus-like M2-brane configurations, so that 
the world-volume coordinates $(\xi^0,\xi^1,\xi^2)$ are chosen as $(t,\phi,\chi)$. 
Then we assume that the scalar fields $r$ and $\theta$ are functions
of $t$, and $x^{11}$ depend on $t$ and $\phi$ as follows:
\begin{alignat}{3}
  x^{11}(t,\phi) = R_{11} (L\phi + \psi(t)).
\end{alignat}
Here $L$ is the winding number.
Thus the M2-brane which we examine in this subsection is extending 
to the $\chi$ direction and to a slanting direction 
in the $(\phi, x^{11})$-plane (Fig.\ref{fig:torus}).
The M2-brane also has momentum along the 11th direction.
By using these settings, the metric induced on the M2-brane 
and the pullback of the 3-form gauge field become
\begin{alignat}{3}
  P[G]_{ab} &= 
  \begin{pmatrix}
    -f^{-\frac{1}{3}}(1 \!-\! f\dot{r}^2 
    \!-\! fr^2\dot{\theta}^2 \!-\! fR_{11}^2 \dot{\psi}^2 ) & 
    f^{\frac{2}{3}} L R_{11}^2 \dot{\psi} & 0 \\
    f^{\frac{2}{3}} L R_{11}^2 \dot{\psi} & f^{\frac{2}{3}} 
    (r^2 \cos^2\theta + L^2 R_{11}^2) & 0 \\
    0 & 0 & f^{\frac{2}{3}} r^2 \sin^2\theta
  \end{pmatrix}, \notag
  \\[0.2cm]
  P[A] &= N \ell_p^3 \sin^2\theta \dot{\psi} \;
  dt \wedge d\phi \wedge d\chi. 
\end{alignat}
The Lagrangian is obtained by inserting the above data
into the action (\ref{eq:M2ac}). It becomes
\begin{alignat}{3}
  \mathcal{L}_{\text{M2}} &= - 4\pi^2 T_2 f^{\frac{1}{2}} r 
  \sqrt{r^2 \cos^2\theta \!+\! L^2 R_{11}^2} \sin\theta
  \sqrt{ 1 \!-\! f\dot{r}^2 \!-\! fr^2\dot{\theta}^2
  - \frac{f r^2 R_{11}^2 \cos\theta^2}{r^2 \cos^2\theta \!+\! L^2 R_{11}^2}
  \dot{\psi}^2} \notag
  \\
  &\qquad + N\sin^2\theta \dot{\psi} .
\end{alignat}
From this Lagrangian, we see that a momentum conjugate to $\psi$
becomes a quantized conserved quantity and the second term yields
a negative potential energy if $\dot{\psi}$ is positive.

In order to examine the potential energy of the 
torus-like M2-brane configurations, 
it is appropriate to move to the Hamiltonian formalism. 
The canonical conjugate momenta are defined as
\begin{alignat}{3}
  P_r &\equiv \frac{\partial \mathcal{L}_{\text{M2}}}{\partial \dot{r}} =
  \frac{4\pi^2 T_2 f^{\frac{3}{2}} r \sqrt{r^2\cos^2\theta
  \!+\! L^2 R_{11}^2} \sin\theta \, \dot{r}}
  { \sqrt{1 \!-\! f\dot{r}^2 \!-\! fr^2\dot{\theta}^2
  - \frac{f r^2 R_{11}^2 \cos\theta^2}{r^2 \cos^2\theta + L^2 R_{11}^2}
  \dot{\psi}^2} }, 
  \notag
  \\
  P_\theta &\equiv \frac{\partial \mathcal{L}_{\text{M2}}}
  {\partial \dot{\theta}} =
  \frac{4\pi^2 T_2 f^{\frac{3}{2}} r^3 \sqrt{r^2\cos^2\theta
  \!+\! L^2 R_{11}^2} \sin\theta \, \dot{\theta}} 
  { \sqrt{1 \!-\! f\dot{r}^2 \!-\! fr^2\dot{\theta}^2
  - \frac{f r^2 R_{11}^2 \cos\theta^2}{r^2 \cos^2\theta + L^2 R_{11}^2}
  \dot{\psi}^2} }, 
  \\
  P_\psi &\equiv \frac{\partial \mathcal{L}_{\text{M2}}}{\partial \dot{\psi}} = 
  \frac{4\pi^2 T_2 f^{\frac{3}{2}} r^3 (r^2\cos^2\theta \!+\! 
  L^2 R_{11}^2)^{-\frac{1}{2}} R_{11}^2 \sin\theta \cos^2\theta \, \dot{\psi}}
  { \sqrt{1 \!-\! f\dot{r}^2 \!-\! fr^2\dot{\theta}^2
  - \frac{f r^2 R_{11}^2 \cos\theta^2}{r^2 \cos^2\theta + L^2 R_{11}^2}
  \dot{\psi}^2} } + N\sin^2\theta,
  \notag
\end{alignat}
and the Hamiltonian is given by
\begin{alignat}{3}
  \mathcal{H}_{\text{M2}} &= f^{-\frac{1}{6}}
  \bigg[ \bigg( \frac{P_r}{f^{\frac{1}{3}}} \bigg)^2 + 
  \bigg(\frac{P_\theta}{f^{\frac{1}{3}}r} \bigg)^2 + 
  \bigg( \frac{P_\psi \!-\! N\sin^2\theta}{f^{\frac{1}{3}}r R_{11} \cos\theta
  (r^2\cos^2\theta \!+\! L^2 R_{11}^2)^{-\frac{1}{2}}} \bigg)^2 \notag
  \\ 
  &\qquad\qquad + \Big( 4\pi^2 T_2 f^{\frac{2}{3}} r \sqrt{r^2\cos^2\theta
  \!+\! L^2 R_{11}^2} \sin\theta \Big)^2 \bigg]^{\frac{1}{2}}
  \label{eq:stHam}
  \\
  &= \sqrt{\frac{P_r^2}{f} + \frac{P_\theta^2}{fr^2}
  + \frac{r^2\cos^2\theta + L^2R_{11}^2}{fr^2R_{11}^2 \cos^2\theta}
  (P_\psi^2 - 2NP_\psi \sin^2\theta + N^2 \sin^2\theta) }. \notag
\end{alignat}
Physical meanings of the first and the second lines are as follows.
As before, the coefficient of the square root is the red shift factor, and
the first and second terms in the square root are kinetic energies
of the $r$ and $\theta$ directions respectively.
The third term in the square root corresponds to the momentum conjugate to the
$\psi$ direction reduced by the effect of background flux of M5-branes. 
The fourth term represents the mass of the torus-like M2-brane.
Especially, the value
$f^{\frac{1}{3}} \sqrt{r^2\cos^2\theta \!+\! L^2 R_{11}^2}$
is the length of the M2-brane extending in the $(\phi,x^{11})$-plane.

Now let us investigate dynamical aspects of this system.
Equations of motion are given by
\begin{alignat}{3}
  &\dot{r} = \frac{P_r}{f \mathcal{H}_{\text{M2}}}, \qquad
  \dot{P_r} = -\frac{P_r^2}{fr \mathcal{H}_{\text{M2}}}
  -\frac{1}{frR_{11}^2 \mathcal{H}_{\text{M2}}}
  \big( P_\psi^2 \!-\! 2NP_\psi \sin^2\theta 
  \!+\! N^2\sin^2\theta \big), \notag
  \\[0.1cm]
  &\dot{\theta} = \frac{P_\theta}{fr^2 \mathcal{H}_{\text{M2}}}, \qquad
  \dot{P_\theta} = - \frac{L^2\sin\theta}
  {fr^2 \cos^3\theta \mathcal{H}_{\text{M2}}}
  \bigg[(P_\psi \!-\! N)^2 + \frac{N(N \!-\! 2P_\psi)r^2}
  {L^2 R_{11}^2}\cos^4\theta \bigg] , \label{eq:eom}
  \\[0.1cm]
  &\dot{\psi} = \frac{r^2\cos^2\theta \!+\! L^2 R_{11}^2}
  {fr^2 R_{11}^2 \cos^2\theta \, \mathcal{H}_{\text{M2}}} 
  (P_\psi \!-\! N\sin^2\theta), \qquad \dot{P_\psi} = 0. \notag
\end{alignat}
$P_\psi$ takes an integer value.
In order to obtain a static solution, we put $\dot{\theta}=0$.
This condition means $P_{\theta}=0$, and hence $\dot{P_{\theta}}=0$.
Then a singular solution of the form
\begin{alignat}{3}
  P_\psi = N, \quad \theta = \frac{\pi}{2}, \quad
  r = \text{const}, \quad \mathcal{H}_{\text{M2}} = \frac{LN}{f^{\frac{1}{2}}r},
\end{alignat}
can be obtained. Note that, to obtain the above Hamiltonian, we should 
first set $P_\psi = N$ and next substitute $\theta=\frac{\pi}{2}$.
This solution represents the cylindrical M2-brane extending along
the 11th direction, and the same as the singular
solution obtained in the previous subsection.
\begin{figure}[b]
  \begin{center}
    \includegraphics[width=14cm,height=8cm,keepaspectratio]{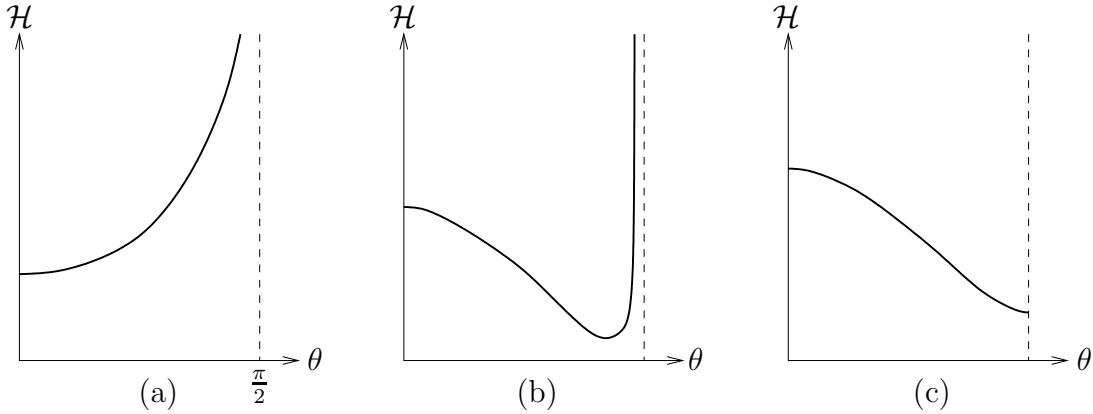}
  \begin{picture}(410,0)
    \put(2,142){$\mathcal{H}$}
    \put(116,12){$\theta$}
    \put(94,4){$\frac{\pi}{2}$}
    \put(52,0){(a)}
    \put(147,142){$\mathcal{H}$}
    \put(260,12){$\theta$}
    \put(195,0){(b)}
    \put(292,142){$\mathcal{H}$}
    \put(407,12){$\theta$}
    \put(345,0){(c)}
  \end{picture}
  \caption{(a) corresponds to the case of $0 \leq P_\psi < \frac{N}{2}$.
  (b) does to that of $\frac{N}{2} < P_\psi$, except for
  $P_\psi=N$ which is drawn in (c).}
  \label{fig:fig2}
  \end{center}
\end{figure}

In the remainder of this subsection, we estimate some dynamical aspects
of the torus-like M2-brane for an arbitrary value of $P_\psi$.
It is simple to choose $P_r=P_\theta=0$ as initial conditions at $t=0$.
Then the Hamiltonian (\ref{eq:stHam}) becomes of the form
\begin{alignat}{3}
  \mathcal{H}_{\text{M2}} &= \sqrt{ \bigg( \frac{1}{fR_{11}^2} + 
  \frac{L^2}{fr^2 \cos^2\theta} \bigg)
  \big(P_\psi^2 - 2NP_\psi \sin^2\theta + N^2\sin^2\theta \big) }, 
  \label{eq:initHam}
\end{alignat}
where $r$ should be considered as some initial value.
The partial differential on $\theta$ is given by
\begin{alignat}{3}
  \frac{\partial \mathcal{H}_{\text{M2}}}{\partial \theta} &=
  \frac{L^2\sin\theta}{fr^2 \cos^3\theta \, \mathcal{H}_{\text{M2}}}
  \bigg[(P_\psi \!-\! N)^2 + \frac{N(N \!-\! 2P_\psi)r^2}
  {L^2 R_{11}^2}\cos^4\theta \bigg],
\end{alignat}
and this equation becomes zero when $\theta=0$ or
\begin{alignat}{3}
  \cos^2\theta = \frac{LR_{11}|P_\psi-N|}{r\sqrt{N(2 P_\psi - N)}},
  \label{eq:theta}
\end{alignat} 
for $\frac{N}{2} < P_\psi$. For the latter case, 
$0 \leq \cos^2\theta \leq 1$ requires that it exists
in the range of
\begin{alignat}{3}
  &\frac{Nr^2}{L^2R_{11}^2} \sqrt{1 \!+\! \frac{L^2R_{11}^2}{r^2}}
  \bigg( \sqrt{1 \!+\! \frac{L^2R_{11}^2}{r^2}} - 1 \bigg) \leq P_{\psi}
  \leq \frac{Nr^2}{L^2R_{11}^2} \sqrt{1 \!+\! \frac{L^2R_{11}^2}{r^2}}
  \bigg( \sqrt{1 \!+\! \frac{L^2R_{11}^2}{r^2}} + 1 \bigg).
\end{alignat}
As mentioned in the subsection \ref{sec:2-1}, our analyses are valid when
$1 \ll \sqrt{N} \ll r/R_{11}$, so the above restriction can be
approximated as
\begin{alignat}{3}
  \frac{N}{2} < P_\psi < \frac{2N r^2}{L^2R_{11}^2} \sim \infty. 
\end{alignat}

When $P_\psi \leq \frac{N}{2}$, a global minimum of the Hamiltonian
(\ref{eq:initHam}) exists at $\theta=0$. In this case,
from the equations of motion (\ref{eq:eom}), we see that $r$
decreases and $\theta$ remains zero as time evolves.
When $\frac{N}{2} < P_\psi$, the global minimum exists 
at the $\theta$ given by the equation (\ref{eq:theta}), and we see
that both $r$ and $\theta$ decrease as time evolves. As explained before,
the singular case occurs when $P_\psi = N$ (Fig.\ref{fig:fig2}).

\subsection{Torus-like M2-brane revisited from D2-brane action} \label{sec:3-3}

In the previous subsection, the configurations of the torus-like M2-brane with
winding and momentum numbers of the 11th direction are discussed.
Such discussions can also be made by employing
the D2-brane action. The winding number and momentum along the 11th
direction correspond
to the electric and magnetic flux on the D2-brane world-volume, respectively.
In this subsection we reproduce the Hamiltonian (\ref{eq:stHam})
from the Born-Infeld action for the D2-brane.

The Born-Infeld action for a single D2-brane is given by the form
\begin{alignat}{3}
  S_{\text{D2}} &= -T_2 \int d^3\xi \, e^{-\phi}
  \sqrt{-\det (P[G]_{ab} + P[B]_{ab} + \lambda F_{ab})}. \label{eq:actD2}
\end{alignat}
Here $\lambda \equiv 2\pi \ell_s^2$ and $F_{ab}$ is a gauge field strength
on the D2-brane. 
Now we will construct the same situation as that of the previous subsection.
The space-time coordinates are chosen as the equations (\ref{eq:S3}).
Then the background metric, dilaton field and the NS-NS 2-form 
of the equations (\ref{eq:NS5}) become of the forms
\begin{alignat}{3}
  ds_{10}^2 &= \eta_{\mu\nu}dx^\mu dx^\nu + f(dr^2 + r^2 d\theta^2
  + r^2 \cos^2\theta d\phi^2 + r^2 \sin^2\theta d\chi^2), \quad
  e^\phi = f^{\frac{1}{2}}, 
  \\
  B &= N \ell_s^2 \sin^2\theta d\phi \wedge d\chi. \notag
\end{alignat}
We choose the world-volume coordinates $(\xi^0,\xi^1,\xi^2)$ on the D2-brane
as $(t,\phi,\chi)$, and suppose that $r$ and $\theta$ are functions of $t$.
As for the gauge field strength, we assume that $F_{t\chi}$ and $F_{\chi\phi}$
are the only non zero components which only depend on $t$. By using these setups,
the $3\times 3$ matrix in the square root of the action (\ref{eq:actD2})
becomes
\begin{alignat}{3}
  P[G]_{ab} \!+\! P[B]_{ab} \!+\! \lambda F_{ab} =\! 
  \begin{pmatrix}
    -(1 \!-\! f\dot{r}^2 \!-\! N\ell_s^2 \dot{\theta}^2) & 0 & \lambda F_{t\chi}
    \\
    0 & 
    N\ell_s^2 \cos^2\theta & N\ell_s^2 \sin^2\theta \!-\! \lambda F_{\chi\phi}
    \\
    -\lambda F_{t\chi} & - N\ell_s^2 \sin^2\theta \!+\! \lambda F_{\chi\phi} & 
    N\ell_s^2 \sin^2\theta
  \end{pmatrix},
\end{alignat}
and the Lagrangian is expressed as
\begin{alignat}{3}
  \mathcal{L}_{\text{D2}} &= - 4\pi^2 T_2 e^{-\phi}
  \sqrt{ X(1 \!-\! f\dot{r}^2 \!-\! N\ell_s^2 \dot{\theta}^2) - 
  N\ell_s^2 \cos^2\theta (\lambda F_{t\chi})^2}, \label{eq:lagD2}
  \\
  &\quad X \equiv N^2\ell_s^4\sin^2\theta - 2N\ell_s^2\sin^2\theta 
  (\lambda F_{\chi\phi}) + (\lambda F_{\chi\phi})^2 . \notag
\end{alignat}

We will deform this Lagrangian using the methods of 
refs.\cite{Emp,PTMM,PTMM2,Hya}.
First, from the magnetic flux quantization condition
\begin{alignat}{3}
  \int d\chi d\phi F_{\chi\phi} = 2\pi M,
\end{alignat}
we obtain the relation $F_{\chi\phi} = M/2\pi$. $M$ takes an integer value.
Then the Lagrangian (\ref{eq:lagD2}) is written as
\begin{alignat}{3}
  \mathcal{L}_{\text{D2}} &= - 4\pi^2 T_2 e^{-\phi}
  \sqrt{ X(1 \!-\! f\dot{r}^2 \!-\! N\ell_s^2 \dot{\theta}^2) - 
  N\ell_s^2 \cos^2\theta (\lambda F_{t\chi})^2}, \label{eq:actlag}
  \\
  &\quad X = N^2\ell_s^4\sin^2\theta - 2NM\ell_s^4\sin^2\theta + M^2 \ell_s^4 .
  \notag
\end{alignat}

Second, let us define 
\begin{alignat}{3}
  E \equiv \frac{\partial \mathcal{L}_{\text{D2}}}{\partial (\lambda F_{t\chi})}
  = \frac{4\pi^2 T_2 e^{-\phi}N\ell_s^2 \cos^2\theta (\lambda F_{t\chi})}
  {\sqrt{X(1 \!-\! f\dot{r}^2 \!-\! N\ell_s^2 \dot{\theta}^2) - 
  N\ell_s^2 \cos^2\theta (\lambda F_{t\chi})^2}}.
\end{alignat}
$E$ can be just a constant by employing an equation of motion for $A_\chi$.
And from this equation $\lambda F_{t\chi}$ is expressed as
\begin{alignat}{3}
  \lambda F_{t\chi} &= \frac{E \sqrt{X(1 \!-\! f\dot{r}^2 \!-\! 
  N\ell_s^2 \dot{\theta}^2) }}{\sqrt{N}\ell_s \cos\theta
  \sqrt{E^2 + (4\pi^2 T_2 e^{-\phi})^2 N\ell_s^2 \cos^2\theta}}.
\end{alignat}
The electric flux quantization condition is given by the form
\begin{alignat}{3}
  E &= 2\pi T L = \frac{L}{\ell_s^2},
\end{alignat}
where $L$ is an integer.
Now we make a Legendre transformation in the following way.
\begin{alignat}{3}
  \mathcal{\tilde{L}}_{\text{D2}} &\equiv \mathcal{L}_{\text{D2}} 
  - E (\lambda F_{t\chi}) \notag
  \\
  &= - \frac{\sqrt{E^2 + (4\pi^2 T_2 e^{-\phi})^2 N\ell_s^2 \cos^2\theta}}
  {\sqrt{N}\ell_s \cos\theta}
  \sqrt{X(1 \!-\! f\dot{r}^2 \!-\! 
  N\ell_s^2 \dot{\theta}^2) }.
\end{alignat}
Note that this does not coincide with the Lagrangian obtained in 
the previous subsection.

As usual we move to the Hamiltonian formalism.
The canonical conjugate momenta are defined as
\begin{alignat}{3}
  P_r &\equiv \frac{\partial \mathcal{\tilde{L}}_{\text{D2}}}{\partial \dot{r}}
  = \frac{\sqrt{E^2 + (4\pi^2 T_2 e^{-\phi})^2 N\ell_s^2 \cos^2\theta}}
  {\sqrt{N}\ell_s \cos\theta} \frac{\sqrt{X}f\dot{r}}
  {\sqrt{1 \!-\! f\dot{r}^2 \!-\! N\ell_s^2 \dot{\theta}^2}},
  \\
  P_\theta &\equiv \frac{\partial \mathcal{\tilde{L}}_{\text{D2}}}
  {\partial \dot{\theta}}
  = \frac{\sqrt{E^2 + (4\pi^2 T_2 e^{-\phi})^2 N\ell_s^2 \cos^2\theta}}
  {\sqrt{N}\ell_s \cos\theta} \frac{\sqrt{X}N\ell_s^2 \dot{\theta}}
  {\sqrt{1 \!-\! f\dot{r}^2 \!-\! N\ell_s^2 \dot{\theta}^2}}, \notag
\end{alignat}
and the Hamiltonian is given by
\begin{alignat}{3}
  \mathcal{H}_{\text{D2}} &= \sqrt{ \frac{P_r^2}{f} + \frac{P_\theta^2}{fr^2}
  + \frac{L^2 R_{11}^2 \!+\! r^2\cos^2\theta}{fr^2R_{11}^2 \cos^2\theta}
  (N^2\sin^2\theta \!-\! 2NM\sin^2\theta \!+\! M^2) }.
\end{alignat}
Thus we have reproduced the Hamiltonian (\ref{eq:stHam}).
The description of the D2-brane agrees with
that of the M2-brane completely at the level of the Hamiltonian.


Before ending this subsection, it is meaningful to expand 
the Lagrangian (\ref{eq:actlag}) so as to neglect the higher order terms
of $\frac{N}{M}$. It becomes as follows:
\begin{alignat}{3}
  \mathcal{L}_{\text{D2}} &= -M T_0 e^{-\phi}
  \sqrt{1 \!-\! f\dot{r}^2 \!-\! N\ell_s^2 \dot{\theta}^2} 
  \bigg( 1 - \frac{N}{M}\sin^2\theta +
  \frac{N^2}{2M^2}\sin^2\theta\cos^2\theta \label{eq:actlag2}
\\
  &\qquad\qquad\qquad\qquad
  - \frac{N\cos^2\theta}{2M^2\ell_s^2(1 \!-\! f\dot{r}^2 \!-\! N\ell_s^2 \dot{\theta}^2)}
  (\lambda F_{t\chi})^2 + \mathcal{O}\big((\tfrac{N}{M})^3\big) \bigg). \notag
\end{alignat}
If we trace the same process as before in this subsection, 
the electric flux quantization condition is given by
\begin{alignat}{3}
  E &= \frac{NT_0 e^{-\phi} \cos^2\theta (\lambda F_{t\chi})}
  {M\ell_s^2 \sqrt{1 \!-\! f\dot{r}^2 \!-\! N\ell_s^2 \dot{\theta}^2}}
  = \frac{L}{\ell_s^2}, \label{eq:eleflux}
\end{alignat}
and the Hamiltonian becomes
\begin{alignat}{3}
  \mathcal{H} &= \sqrt{\frac{P_r^2}{f} + \frac{P_\theta^2}{N\ell_s^2}
  + M^2T_0^2 e^{-2\phi} \bigg( 1 \!-\! \frac{N}{M}\sin^2\theta 
  \!+\! \frac{N^2}{2M^2}\sin^2\theta\cos^2\theta 
  + \frac{R_{11}^2 L^2}{2r^2\cos^2\theta} \bigg)^2 }.
\end{alignat}
Let us roughly estimate physical features obtained from
this Hamiltonian by setting $P_r = P_\theta = 0$. 
It becomes of the form
\begin{alignat}{3}
  \mathcal{H} &\sim
  M T_0 e^{-\phi} \bigg( 1 \!-\! \frac{N}{M}\sin^2\theta 
  \!+\! \frac{N^2}{2M^2}\sin^2\theta\cos^2\theta 
  + \frac{R_{11}^2 L^2}{2r^2\cos^2\theta} \bigg).
\end{alignat}
The first term is just the mass of $M$ D0-branes.
The second term reduces the potential energy by the effect of
the background flux of NS5-branes .
The third term increases the potential energy,
but the contribution is less than that of the second term
when we compare their absolute values.
If $L=0$, the energy reaches to a minimum
when $\theta=\frac{\pi}{2}$. The existence of the fourth term, however,
shift the minimum of the potential slightly below $\frac{\pi}{2}$.
After all, we obtain toroidal configurations of the D2-brane 
expanding into the $\phi$ and $\chi$ directions, 
with the electric flux along the $\chi$ direction 
and the magnetic flux on it. The radius of the $\phi$ direction
is small compared with that of the $\chi$ direction.
Thus the torus looks almost like a closed string.

\section{Construction of Expanded Strings from D0-branes} \label{sec:D0}

In this section we reexamine the torus-like configurations of the 
M2-brane or D2-brane from the viewpoint of D0-branes.
The point is to realize fundamental strings by only using
the degrees of freedom of D0-branes.
In the BFSS matrix theory, which is defined in the flat space-time, 
fundamental strings have been constructed through a guide of the BPS 
condition\cite{Ima}.
In our case, however, the background is curved and the BPS condition
is also useless.
We use the non-abelian Born-Infeld action for $M$ D0-branes
and aim for reproducing the Lagrangian (\ref{eq:actlag}) or (\ref{eq:actlag2}).

The action for $N$ coincident D-branes for flat space-time has been
constructed in ref.\cite{Tse} and extended in a way consistent 
with T-duality in ref.\cite{Mye}.
The non-abelian Born-Infeld action for $M$ coincident D0-branes 
is given by the form
\begin{alignat}{3}
  S_{M \text{D0}} &= - T_0 \int dt\,
  \text{STr} \Big( e^{-\phi} \sqrt{- \big( P[E]_{00} + P[E']_{00} \big) 
  \det ({Q^i}_j)} \, \Big), \quad i,j=1,\cdots,9. \label{eq:actD0}
\end{alignat}
Here we defined $E_{\mu\nu} = G_{\mu\nu} \!+\! B_{\mu\nu}$,
$E'_{\mu\nu} = E_{\mu i}({Q^{-1 i}}_j-\delta^i_j)E^{jk}E_{k\nu}$
and ${Q^i}_j = \delta^i_j \!+\! \frac{i}{\lambda}[X^i,X^k]E_{kj}$.
$X^i$ belong to the adjoint representation of $U(M)$ and 
denote the positions of $M$ D0-branes.
The partial differentials appearing in the pullbacks
should be regarded as covariant derivatives
with respect to gauge symmetry on the world-volume.
The background metric is given by the equation (\ref{eq:NS5}),
and the NS-NS 2-form is written as
\begin{alignat}{3}
  B &= \frac{f}{(x^6)^2 \!+\! (x^7)^2}
  \big( x^7 x^9 dx^6 \wedge dx^8 - x^7 x^8 dx^6 \wedge dx^9 
\\
  &\qquad\qquad\qquad
  - x^6 x^9 dx^7 \wedge dx^8 + x^6 x^8 dx^7 \wedge dx^9 \big) , \notag
\end{alignat}
where the gauge is fixed as in the previous subsection.

In the subsection \ref{sec:3-2} and \ref{sec:3-3}, 
we have investigated the torus-like configurations of M2-brane or D2-brane
expanding into the $S^3$. In the matrix theory, the toroidal configuration
can be realized by using the matrices called `shift' and `clock'.
In our case, what we want to realize is the torus configuration
whose two 1-cycles become $S^1$s in the $(x^6,x^7)$-plane and $(x^8,x^9)$-plane
respectively.
Then it is natural to assume that the matrices $X^i$ become
$X^1 = \cdots = X^5 = 0$ and
\begin{alignat}{3}
  &X^6 = \frac{r\cos\theta}{2}
  \begin{pmatrix}
    0 & 1 &   &      &   &   & 1  \\
    1 & 0 & 1 &      &   &   &    \\
      & 1 & 0 &      &   &   &    \\
      &   &   &\ddots&   &   &    \\
      &   &   &      & 0 & 1 &    \\
      &   &   &      & 1 & 0 & 1  \\
    1 &   &   &      &   & 1 & 0
  \end{pmatrix} ,\quad
  X^7 = \frac{r\cos\theta}{2}
  \begin{pmatrix}
    0 & -i &   &      &   &   & i  \\
    i & 0  & -i&      &   &   &    \\
      & i  & 0 &      &   &   &    \\
      &    &   &\ddots&   &   &    \\
      &    &   &      & 0 & -i&    \\
      &    &   &      & i & 0 & -i \\
    -i&    &   &      &   & i & 0
  \end{pmatrix}, \notag
  \\
  &X^8 = r\sin\theta
  \begin{pmatrix}
    c_1&    &      &        &     \\
       & c_2&      &        &     \\
       &    &\ddots&        &     \\
       &    &      & c_{M-1}&     \\
       &    &      &        & c_M
  \end{pmatrix}, \quad
  X^9 = r\sin\theta
  \begin{pmatrix}
    s_1&    &      &        &     \\
       & s_2&      &        &     \\
       &    &\ddots&        &     \\
       &    &      & s_{M-1}&     \\
       &    &      &        & s_M 
  \end{pmatrix}.
\end{alignat}
Here $c_m \equiv \cos\frac{2\pi m}{M}$ and $s_m \equiv \sin\frac{2\pi m}{M}$ 
are defined for abbreviations.
Note that the relations 
$(X^6)^2 + (X^7)^2 \!=\! r^2 \cos^2\theta \mathbf{1}_M$
and $(X^8)^2 + (X^9)^2 \!=\! r^2 \sin^2\theta \mathbf{1}_M$ hold.
$\mathbf{1}_{M}$ denotes the $M \!\times \!M$ identity matrix.
The commutation or anti-commutation relations of these matrices 
or other useful relations are collected in the appendix \ref{app:1}.

Now we substitute these expressions into the action (\ref{eq:actD0}).
There are a few remarks on practicing calculations.
First, each component of the NS-NS 2-form becomes an $M \!\times\! M$ matrix.
In order to be Hermite matrices, they should become of the forms
\begin{alignat}{3}
  B_{68} &= \frac{f\{X^7,X^9\}}{2r^2\cos^2\theta} , \quad
  B_{69} = - \frac{f\{X^7,X^8\}}{2r^2\cos^2\theta} , \quad
  \\
  B_{78} &= - \frac{f\{X^6,X^9\}}{2r^2\cos^2\theta} , \quad
  B_{79} = \frac{f\{X^6,X^8\}}{2r^2\cos^2\theta},\notag
\end{alignat}
where we introduced the notation $\{X^i,X^j\} \equiv X^iX^j + X^jX^i$.
Next, because of the symmetrized trace operation in the action (\ref{eq:actD0}), 
$D_0 X^i$, $[X^i,X^j]$ and $B_{ij}$ should be ordered symmetrically.
Third, we expand the action to the order of $f^2[A_0,X^i]^2$ or $f^2[X^i,X^j]^2$.
From explicit calculations, we see that
the term $f[X^6,X^8]$ is of the order of $\frac{N}{M}$, for example.
So we omit above the order of $(\frac{N}{M})^2$
in the action. 

At the beginning, by taking care of the above remarks, we obtain
\begin{alignat}{3}
  - \big( P[E]_{00} + P[E']_{00} \big) &= \mathbf{1}_{M} 
  - D_0 X^i D_0 X^j G_{ik} {Q^{-1 k}}_j \notag
  \\
  &= \big( 1 - f\dot{r}^2 - N\ell_s^2 \dot{\theta}^2 \big) \mathbf{1}_{M}
  + f[A_0,X^i]^2 + \mathcal{O} \big((\tfrac{N}{M})^3 \big). \label{eq:E00}
\end{alignat}
Now $i$ takes only $6,7,8,9$.
Next let us define ${S^i}_j \equiv \frac{i}{\lambda}[X^i,X^k]E_{kj}$. Then
we can obtain a matrix valued equation of the form
\begin{alignat}{3}
  \det {Q^i}_j = \mathbf{1}_M + \text{Tr}_i S  + \tfrac{1}{2} 
  \big(\text{Tr}_iS \big)^2 - \tfrac{1}{2} \text{Tr}_i S^2 + \mathcal{O}(S^3).
  \label{eq:Q}
\end{alignat}
The notation Tr$_i$ indicates that the trace should be taken with
respect to the indices $i$.
After some calculations which is explained in detail in the appendix \ref{app:1},
the equation (\ref{eq:Q}) is translated into the form
\begin{alignat}{3}
  \det {Q^i}_j = \Big(1 - \frac{2N}{M}\sin^2\theta 
  + \frac{N^2}{M^2}\sin^2\theta \Big) \mathbf{1}_M
  + \mathcal{O}\big((\tfrac{N}{M})^3 \big). \label{eq:detQ}
\end{alignat}
By substituting the equations (\ref{eq:E00}) and (\ref{eq:detQ})
into the action (\ref{eq:actD0}), the Lagrangian becomes
\begin{alignat}{3}
  \mathcal{L}_{M\text{D0}} &= -T_0 e^{-\phi} \, \text{Tr}
  \Bigg( \sqrt{\frac{X}{M^2\ell_s^4} \big( 1 - f\dot{r}^2 - N\ell_s^2 
  \dot{\theta}^2 \big) \mathbf{1}_{M} + f[A_0,X^i]^2 + 
  \mathcal{O}\big((\tfrac{N}{M})^3 \big) } \, \Bigg) \notag
  \\
  &= -T_0 e^{-\phi} \sqrt{1 \!-\! f\dot{r}^2 \!-\! N\ell_s^2 \dot{\theta}^2} \;
  \text{Tr} \bigg( \Big( 1 \!-\! \frac{N}{M}\sin^2\theta \!+\! \frac{N^2}{2M^2}
  \sin^2\theta\cos^2\theta \Big) \mathbf{1}_{M} \notag
\\
  &\qquad\qquad\qquad
  + \frac{f[A_0,X^i]^2}{2(1 \!-\! f\dot{r}^2 \!-\! N\ell_s^2 \dot{\theta}^2)} 
  + \mathcal{O}\big((\tfrac{N}{M})^3 \big) \bigg). \label{eq:prelag}
\end{alignat}
Here $X$ is already appeared in the equation (\ref{eq:actlag}).

The remaining task is to give an explicit expression for $A_0$.
In order to do this, we should solve the equation
which is obtained by varying $A_0$.
It is given by the form
\begin{alignat}{3}
  [X^i,[A_0,X^i]] = 0. \label{eq:A0}
\end{alignat}
A trivial solution is $A_0=0$, but this is not appropriate to reproduce 
the Lagrangian (\ref{eq:actlag2}). 

To get the expression for $A_0$ which reproduce the Lagrangian 
(\ref{eq:actlag2}), it is useful to consider
the quantization condition of fundamental string charge.
The minimal coupling term between NS-NS 2-form field and
$L$ fundamental strings extending to $x^j$ direction is given by 
\begin{alignat}{3}
  LT \int dt dx^j B_{0j}. \label{eq:quastr}
\end{alignat}
And the matrix realization of this term can be done as follows.
What we want is a minimal coupling term to $B_{0j}$ in the action 
(\ref{eq:actD0}), and it arises from the terms
in the $P[E']_{00}$ of the form
\begin{alignat}{3}
  D_0X^i E'_{i0} + D_0X^i E'_{0i} &\sim \frac{2if}{\lambda}
  D_0X^i [X^i,X^j]B_{0j}.
\end{alignat}
Then the minimal coupling term is given by
\begin{alignat}{3}
  T \int dt \, \frac{f T_0 e^{-\phi}}
  {\sqrt{1 \!-\! f\dot{r}^2 \!-\! N\ell_s^2 \dot{\theta}^2}}
  \text{Tr} \Big( [X^j,X^i][A_0,X^i] \Big) \;B_{0j} \label{eq:min}
\end{alignat}
If $A_0=0$, this term vanishes, but otherwise it yields the non-zero
string charge.

Now we want to realize the
case where the flux corresponding to the fundamental string is 
along the $S^1$ in the $(x^8,x^9)$ plane. Therefore it is reasonable
to assume that $A_0$ is a diagonal matrix,
since $[A_0,X^8]=[A_0,X^9]=0$ and coupling terms of $B_{06}$ and
$B_{07}$ do not appear in this case.
Then, in addition to the trivial one, 
we can obtain a solution of the equation (\ref{eq:A0}):
\begin{alignat}{3}
  a_m - a_{m+1} = \frac{a}{M\ell_s^2} + \mathcal{O}\big((\tfrac{N}{M})^2 \big),
  \qquad m = 1,\cdots,M,
\end{alignat}
where $a_m$ denotes the $m$th diagonal element of $A_0$ and $a_{M+1} = a_1$.
$a$ is a function of $t$.
By inserting this solution into the equation (\ref{eq:min})
and taking the limit of large $M$, the trace is 
substituted for an integral and we obtain
\begin{alignat}{3}
  &T \int dt d\chi \frac{NT_0 e^{-\phi}\cos^2\theta \, a}
  {M \sqrt{1 \!-\! f\dot{r}^2 \!-\! N\ell_s^2 \dot{\theta}^2}}
  \Big( -r\sin\theta\sin\chi \, B_{08} 
  + r \sin\theta\cos\chi \, B_{09} \Big) \notag
  \\
  =\; &T \int dt d\chi \frac{NT_0 e^{-\phi}\cos^2\theta \, a}
  {M \sqrt{1 \!-\! f\dot{r}^2 \!-\! N\ell_s^2 \dot{\theta}^2}} B_{t\chi},
\end{alignat}
where $\chi$ is identified with the angular coordinate
in the $(x^8,x^9)$-plane, as like in the previous section.
Comparing this equation with (\ref{eq:quastr}), the following
quantization condition
\begin{alignat}{3}
  \frac{NT_0 e^{-\phi}\cos^2\theta \, a}
  {M \sqrt{1 \!-\! f\dot{r}^2 \!-\! N\ell_s^2 \dot{\theta}^2}} &= L,
\end{alignat}
can be obtained.
This is similar to the equation (\ref{eq:eleflux}) given in the previous
subsection, and becomes equal if we make the following identification
\begin{alignat}{3}
  a = \lambda F_{t\chi}.
\end{alignat}

Then straightforward calculation leads
\begin{alignat}{3}
  f[A_0,X^i]^2 &= - \frac{N}{M^2\ell_s^2} \cos^2\theta \, (\lambda F_{t\chi})^2
  \, \mathbf{1}_M + \mathcal{O}\big((\tfrac{N}{M})^3\big).
\end{alignat}
By substituting this relation for the Lagrangian (\ref{eq:prelag}), 
we can finally obtain the same Lagrangian as 
(\ref{eq:actlag}) or (\ref{eq:actlag2}).
This means that the dynamics observed in the previous subsections
by using the action of M2-brane or D2-brane 
can be reproduced exactly from the viewpoint of D0-branes.

\vspace{0.5cm}
\section{Conclusions and Discussions} \label{sec:Condis}

In this paper we have discussed the configurations of the spherical, 
cylindrical and torus-like M2-brane in the background of
the infinite array of M5-branes in the $x^{11}$ direction. 
Reducing along the $x^{11}$ direction, this background is interpreted
as that of NS5-branes.
In the region where our calculations
are reliable, all configurations are stabilized against collapse
by virtue of the background flux of M5-branes. 
Through the analyses of those configurations, we have investigated 
the deep relations among M2-brane, D2-brane, fundamental strings and D0-branes.


The spherical M2-brane with momentum along the 11th direction
is identified with the spherical D2-brane with magnetic flux on it.
We reproduced the well known relation
between the radius of the spherical D2-brane and
the number of magnetic flux on it from the viewpoint of M2-brane. 
Moreover we obtained the classical motion for
the radial direction, and find that the spherical D2-brane falls
deep into the throat part of NS5-branes.
Finally the spherical D2-brane will reach to the throat part
of M5-branes and becomes the giant graviton
in the background of $AdS_7 \times S^4$.
We have also checked that the spherical D2-brane configuration is not a BPS state.

The cylindrical M2-brane extending along the 11th direction
is identified with closed strings. This closed strings has the angular
momentum along certain $S^1$ in the $S^3$ where the 
flux of NS5-branes penetrates.
In general closed strings shrinks to zero size because of its tension. 
However, when the number of 
angular momentum is equal to that of the background flux of NS5-branes,
these closed strings can freely 
expand in the $S^3$ without loss of the energy.
Expanded closed strings are considered as 
the counterpart of giant graviton, where the reduction is 
done for one of the world-volume direction of the giant graviton. 
In order to check this correspondence, however, the careful analysis
is needed, because the geometry where the giant graviton lives is $S^4$, and 
on the other hand the background of NS5-branes is $S^3\times S^1$.
There may be a transition from the cylindrical M2-brane to the 
spherical M2-brane. A similar kind of transition is
discussed in the context of the type IIA string theory in ref.\cite{Hya}.

The torus-like M2-brane with winding and momentum numbers along
the 11th direction is identified with the toroidal D2-brane with
electric and magnetic fluxes on it.
In fact, we obtained the same Hamiltonian of the toroidal D2-brane 
as that of the torus-like M2-brane.
We should stress that we could as well
reproduce the same Hamiltonian by beginning with the non-abelian
Born-Infeld action for a large number of D0-branes.
This is just the construction of closed strings
as a bound state of D0-branes.
By using the torus-like configurations, we could have
deep understanding on the relations among M2-brane, D2-brane,
fundamental strings, and D0-branes.

In the cases of spherical and torus-like M2-branes,
they fall deep into the horizon of the infinite array of M5-branes, and 
would finally become the giant gravitons.
The reason for these motions of the radial direction is as follows.
In terms of type IIA superstring theory,
these configurations are bound states of D2-brane and D0-branes, so
their masses depend on the dilaton field.
Their masses become smaller if they approach the origin of the radial direction.
On the other hand, the cylindrical M2-brane is fully static.
This is due to the fact that
the tension of closed string does not depend on the dilaton field.
The throat geometry of NS5-branes is also well described by 
the Wess-Zumino-Witten model on the group manifold of $SU(2)$, 
and it is an interesting problem to realize the cylindrical or
torus-like M2-brane configurations in that model.

In section \ref{sec:D0} the quantization condition of string charge
in the background of NS5-branes
is argued from the viewpoint of D0-branes. The same results are 
also applied to other D$p$-branes for $1\leq p$.
For example, in the background of flat space-time, we obtain
the following effective action:
\begin{alignat}{3}
  S_{M \text{D}p} &\sim -MV_{p+1}T_p
  - T_p \int d^{p+1}\sigma \, \text{Tr}
  \Big( \frac{1}{2}(D_a X^i)(D^a X^i) + \frac{\lambda^2}{4}F^{ab}F_{ab}
  - \frac{1}{4\lambda^2}[X^i,X^j]^2 \Big)
  \\
  &\qquad\qquad\qquad - \frac{iT_p}{\lambda} \int d^{p+1}\sigma \, \text{Tr}
  \Big( [X^i,X^j]D_0 X^j \Big) B_{0i}. \notag
\end{alignat}
The second line in the above equation gives the string charge.
Especially in the case of D0-branes, we obtain the quantization condition
\begin{alignat}{3}
  \frac{i}{R_{11}} \text{Tr}\Big( [X^i,X^j]D_0 X^j \Big) = L,
\end{alignat}
where $L$ takes an integer value. This coincides with the results obtained
from arguments of 11-dimensional superalgebras\cite{BSS}.
Non-trivial $A_0$ have played an important role in ref\cite{BL},
which is an matrix version of ref\cite{MT}.


\vspace{0.5cm}
\section*{acknowledgments}

I would like to thank Kazuo Hosomichi for useful discussions and
careful reading of this manuscript, Takeshi Fukuda and Hiroshi Kunitomo 
for useful discussions and comments.
I also thank Tsuguhiko Asakawa, Masafumi Fukuma, Isao Kishimoto, 
Masao Ninomiya, Masatoshi Nozaki 
and Sachiko Ogushi for discussions or encouragements.

\appendix
\newpage
\section{Calculations} \label{app:1}

The definitions and calculations employed in the section \ref{sec:D0} are
collected in this appendix. 
\begin{alignat}{3}
  &(X^6)^2 + (X^7)^2 = r^2 \cos^2\theta \; \mathbf{1}_{M} ,\quad
  (X^8)^2 + (X^9)^2 = r^2 \sin^2\theta \; \mathbf{1}_{M} , \notag
  \\
  &c^{-}_m \equiv \cos\tfrac{2\pi m}{M} - \cos\tfrac{2\pi (m+1)}{M}, \quad
  s^{-}_m \equiv \sin\tfrac{2\pi m}{M} - \sin\tfrac{2\pi (m+1)}{M}, 
  \\
  &c^{+}_m \equiv \tfrac{1}{2}
  \big(\cos\tfrac{2\pi m}{M} + \cos\tfrac{2\pi (m+1)}{M}\big), \quad
  s^{+}_m \equiv \tfrac{1}{2}
  \big(\sin\tfrac{2\pi m}{M} + \sin\tfrac{2\pi (m+1)}{M}\big). \notag
\end{alignat}
The commutation relations for $X^i$.
\begin{alignat}{3}
  &[X^6,X^7] = [X^8,X^9] = 0, \notag
  \\
  &[X^6,X^8] = \frac{r^2\sin\theta \cos\theta}{2}
  \begin{pmatrix}
       0    &-c^{-}_1&        &      &           &            &c^{-}_{M}    \\
    c^{-}_1 &   0    &-c^{-}_2&      &           &            &             \\
            & c^{-}_2&    0   &      &           &            &             \\
            &        &        &\ddots&           &            &             \\
            &        &        &      &     0     &-c^{-}_{M-2}&             \\
            &        &        &      &c^{-}_{M-2}&     0      &-c^{-}_{M-1} \\
    -c^{-}_M&        &        &      &           & c^{-}_{M-1}&      0 
  \end{pmatrix} ,\notag
  \\
  &[X^6,X^9] = \frac{r^2\sin\theta \cos\theta}{2}
  \begin{pmatrix}
       0    &-s^{-}_1&        &      &           &            &s^{-}_{M}    \\
    s^{-}_1 &   0    &-s^{-}_2&      &           &            &             \\
            & s^{-}_2&    0   &      &           &            &             \\
            &        &        &\ddots&           &            &             \\
            &        &        &      &     0     &-s^{-}_{M-2}&             \\
            &        &        &      &s^{-}_{M-2}&     0      &-s^{-}_{M-1} \\
    -s^{-}_M&        &        &      &           & s^{-}_{M-1}&      0 
  \end{pmatrix} ,
  \\
  &[X^7,X^8] = \frac{ir^2\sin\theta \cos\theta}{2}
  \begin{pmatrix}
       0    & c^{-}_1&        &      &           &            &c^{-}_{M}    \\
    c^{-}_1 &   0    & c^{-}_2&      &           &            &             \\
            & c^{-}_2&    0   &      &           &            &             \\
            &        &        &\ddots&           &            &             \\
            &        &        &      &     0     & c^{-}_{M-2}&             \\
            &        &        &      &c^{-}_{M-2}&     0      & c^{-}_{M-1} \\
     c^{-}_M&        &        &      &           & c^{-}_{M-1}&      0 
  \end{pmatrix} ,\notag
  \\
  &[X^7,X^9] = \frac{ir^2\sin\theta \cos\theta}{2}
  \begin{pmatrix}
       0    & s^{-}_1&        &      &           &            &s^{-}_{M}    \\
    s^{-}_1 &   0    & s^{-}_2&      &           &            &             \\
            & s^{-}_2&    0   &      &           &            &             \\
            &        &        &\ddots&           &            &             \\
            &        &        &      &     0     & s^{-}_{M-2}&             \\
            &        &        &      &s^{-}_{M-2}&     0      & s^{-}_{M-1} \\
     s^{-}_M&        &        &      &           & s^{-}_{M-1}&      0 
  \end{pmatrix} .\notag
\end{alignat}
The anti-commutation relations for $X^i$.
\begin{alignat}{3}
  &\{X^6,X^8\} = r^2\sin\theta \cos\theta
  \begin{pmatrix}
       0    & c^{+}_1&        &      &           &            &c^{+}_{M}    \\
    c^{+}_1 &   0    & c^{+}_2&      &           &            &             \\
            & c^{+}_2&    0   &      &           &            &             \\
            &        &        &\ddots&           &            &             \\
            &        &        &      &     0     & c^{+}_{M-2}&             \\
            &        &        &      &c^{+}_{M-2}&     0      & c^{+}_{M-1} \\
     c^{+}_M&        &        &      &           & c^{+}_{M-1}&      0 
  \end{pmatrix} ,\notag
  \\
  &\{X^6,X^9\} = r^2\sin\theta \cos\theta
  \begin{pmatrix}
       0    & s^{+}_1&        &      &           &            &s^{+}_{M}    \\
    s^{+}_1 &   0    & s^{+}_2&      &           &            &             \\
            & s^{+}_2&    0   &      &           &            &             \\
            &        &        &\ddots&           &            &             \\
            &        &        &      &     0     & s^{+}_{M-2}&             \\
            &        &        &      &s^{+}_{M-2}&     0      & s^{+}_{M-1} \\
     s^{+}_M&        &        &      &           & s^{+}_{M-1}&      0 
  \end{pmatrix} ,
  \\
  &\{X^7,X^8\} = ir^2\sin\theta \cos\theta
  \begin{pmatrix}
       0    &-c^{+}_1&        &      &           &            &c^{+}_{M}    \\
    c^{+}_1 &   0    &-c^{+}_2&      &           &            &             \\
            & c^{+}_2&    0   &      &           &            &             \\
            &        &        &\ddots&           &            &             \\
            &        &        &      &     0     &-c^{+}_{M-2}&             \\
            &        &        &      &c^{+}_{M-2}&     0      &-c^{+}_{M-1} \\
    -c^{+}_M&        &        &      &           & c^{+}_{M-1}&      0 
  \end{pmatrix} ,\notag
  \\
  &\{X^7,X^9\} = ir^2\sin\theta \cos\theta
  \begin{pmatrix}
       0    &-s^{+}_1&        &      &           &            &s^{+}_{M}    \\
    s^{+}_1 &   0    &-s^{+}_2&      &           &            &             \\
            & s^{+}_2&    0   &      &           &            &             \\
            &        &        &\ddots&           &            &             \\
            &        &        &      &     0     &-s^{+}_{M-2}&             \\
            &        &        &      &s^{+}_{M-2}&     0      &-s^{+}_{M-1} \\
    -s^{+}_M&        &        &      &           & s^{+}_{M-1}&      0 
  \end{pmatrix} .\notag
\end{alignat}
Useful relations.
\begin{alignat}{3}
  &c^{-}_m s^{+}_{m+1} - s^{-}_m c^{+}_{m+1} = 
  c^{-}_{m+1} s^{+}_m - s^{-}_{m+1} c^{+}_m =
  \tfrac{1}{2} \sin\tfrac{4\pi}{M} ,
  \\
  &c^{-}_m s^{+}_m - s^{-}_m c^{+}_m = \sin\tfrac{2\pi}{M}. \notag
\end{alignat}
Helpful definitions and calculations.
$[Z^i,Z^j]$ and $\{Z^i,Z^j\}$ are defined as the matrix part of 
$[X^i,X^j]$ and $\{X^i,X^j\}$, respectively.
\begin{alignat}{3}
  &{Y^6}_6 \equiv \frac{i}{2\lambda} \{[X^6,X^8],B_{86}\} 
  + \frac{i}{2\lambda} \{[X^6,X^9],B_{96}\} \notag
  \\
  &\quad\;\; = \frac{N\sin^2\theta}{16\pi} \big( \{[Z^6,Z^8],\{Z^7,Z^9\}\}
  - \{[Z^6,Z^9],\{Z^7,Z^8\}\} \big) \notag
  \\
  &\quad\;\; = \frac{N\sin^2\theta}{16\pi}
  \begin{pmatrix}
    -4\sin\tfrac{2\pi}{M} &0& \sin\tfrac{4\pi}{M} &&& \sin\tfrac{4\pi}{M} &0 \\
    0& -4\sin\tfrac{2\pi}{M} &0&&&& \sin\tfrac{4\pi}{M} \\
    \sin\tfrac{4\pi}{M} &0& -4\sin\tfrac{2\pi}{M} &&&& \\
    &&& \ddots &&& \\
    &&&& -4\sin\tfrac{2\pi}{M} &0& \sin\tfrac{4\pi}{M} \\
    \sin\tfrac{4\pi}{M}&&&&0& -4\sin\tfrac{2\pi}{M} &0 \\
    0& \sin\tfrac{4\pi}{M} &&& \sin\tfrac{4\pi}{M} &0& -4\sin\tfrac{2\pi}{M}
  \end{pmatrix} , \notag
  \\
  &{Y^7}_7 \equiv \frac{i}{2\lambda} \{[X^7,X^8],B_{87}\} 
  + \frac{i}{2\lambda} \{[X^7,X^9],B_{97}\} \notag
  \\
  &\quad\;\; = \frac{N\sin^2\theta}{16\pi} \big(- \{[Z^7,Z^8],\{Z^6,Z^9\}\}
  + \{[Z^7,Z^9],\{Z^6,Z^8\}\} \big) \notag
  \\
  &\quad\;\; = \frac{N\sin^2\theta}{16\pi}
  \begin{pmatrix}
    -4\sin\tfrac{2\pi}{M} &0& -\sin\tfrac{4\pi}{M} &&& -\sin\tfrac{4\pi}{M} &0 \\
    0& -4\sin\tfrac{2\pi}{M} &0&&&& -\sin\tfrac{4\pi}{M} \\
    -\sin\tfrac{4\pi}{M} &0& -4\sin\tfrac{2\pi}{M} &&&& \\
    &&& \ddots &&& \\
    &&&& -4\sin\tfrac{2\pi}{M} &0& -\sin\tfrac{4\pi}{M} \\
    -\sin\tfrac{4\pi}{M}&&&&0& -4\sin\tfrac{2\pi}{M} &0 \\
    0& -\sin\tfrac{4\pi}{M} &&& -\sin\tfrac{4\pi}{M} &0& -4\sin\tfrac{2\pi}{M}
  \end{pmatrix} ,
  \\
  &{Y^8}_8 \equiv \frac{i}{2\lambda} \{[X^8,X^6],B_{68}\} 
  + \frac{i}{2\lambda} \{[X^8,X^7],B_{78}\} \notag
  \\
  &\quad\;\; = \frac{N\sin^2\theta}{16\pi} \big(\{[Z^6,Z^8],\{Z^7,Z^9\}\}
  - \{[Z^7,Z^8],\{Z^6,Z^9\}\} \big) \notag
  \\
  &\quad\;\; = \frac{N\sin^2\theta}{16\pi}
  \begin{pmatrix}
    -4c^{-}_M s^{+}_M -4c^{-}_1 s^{+}_1 &&& \\
    & -4c^{-}_1 s^{+}_1 -4c^{-}_2 s^{+}_2 && \\
    && \ddots && \\
    &&& -4c^{-}_{M-1} s^{+}_{M-1} -4c^{-}_M s^{+}_M
  \end{pmatrix} , \notag
  \\
  &{Y^9}_9 \equiv \frac{i}{2\lambda} \{[X^9,X^6],B_{69}\} 
  + \frac{i}{2\lambda} \{[X^9,X^7],B_{79}\} \notag
  \\
  &\quad\;\; = \frac{N\sin^2\theta}{16\pi} \big(-\{[Z^6,Z^9],\{Z^7,Z^8\}\}
  + \{[Z^7,Z^9],\{Z^6,Z^8\}\} \big) \notag
  \\
  &\quad\;\; = \frac{N\sin^2\theta}{16\pi}
  \begin{pmatrix}
    4s^{-}_M c^{+}_M + 4s^{-}_1 c^{+}_1 &&& \\
    & 4s^{-}_1 c^{+}_1 + 4s^{-}_2 c^{+}_2 && \\
    && \ddots && \\
    &&& 4s^{-}_{M-1} c^{+}_{M-1} + 4s^{-}_M c^{+}_M
  \end{pmatrix} , \notag
\end{alignat}
\begin{alignat}{3}
  &{Y^6}_7 \equiv \frac{i}{2\lambda} \{[X^6,X^8],B_{87}\} 
  + \frac{i}{2\lambda} \{[X^6,X^9],B_{97}\} \notag
  \\
  &\quad\;\; = \frac{iN\sin^2\theta}{16\pi} \big( \{[Z^6,Z^8],\{Z^6,Z^9\}\}
  - \{[Z^6,Z^9],\{Z^6,Z^8\}\} \big) \notag
  \\
  &\quad\;\; = \frac{iN\sin^2\theta}{16\pi}
  \begin{pmatrix}
    0 &0& -\sin\tfrac{4\pi}{M} &&& \sin\tfrac{4\pi}{M} &0 \\
    0& 0 &0&&&& \sin\tfrac{4\pi}{M} \\
    \sin\tfrac{4\pi}{M} &0& 0 &&&& \\
    &&& \ddots &&& \\
    &&&& 0 &0& -\sin\tfrac{4\pi}{M} \\
    -\sin\tfrac{4\pi}{M}&&&&0& 0 &0 \\
    0& -\sin\tfrac{4\pi}{M} &&& \sin\tfrac{4\pi}{M} &0& 0
  \end{pmatrix} , \notag
  \\
  &{Y^7}_6 \equiv \frac{i}{2\lambda} \{[X^7,X^8],B_{86}\} 
  + \frac{i}{2\lambda} \{[X^7,X^9],B_{96}\} \notag
  \\
  &\quad\;\; = \frac{iN\sin^2\theta}{16\pi} \big( \{[Z^7,Z^8],\{Z^7,Z^9\}\}
  - \{[Z^7,Z^9],\{Z^7,Z^8\}\} \big) \notag
  \\
  &\quad\;\; = \frac{iN\sin^2\theta}{16\pi}
  \begin{pmatrix}
    0 &0& -\sin\tfrac{4\pi}{M} &&& \sin\tfrac{4\pi}{M} &0 \\
    0& 0 &0&&&& \sin\tfrac{4\pi}{M} \\
    \sin\tfrac{4\pi}{M} &0& 0 &&&& \\
    &&& \ddots &&& \\
    &&&& 0 &0& -\sin\tfrac{4\pi}{M} \\
    -\sin\tfrac{4\pi}{M}&&&&0& 0 &0 \\
    0& -\sin\tfrac{4\pi}{M} &&& \sin\tfrac{4\pi}{M} &0& 0
  \end{pmatrix} ,
  \\
  &{Y^8}_9 \equiv \frac{i}{2\lambda} \{[X^8,X^6],B_{69}\} 
  + \frac{i}{2\lambda} \{[X^8,X^7],B_{79}\} \notag
  \\
  &\quad\;\; = \frac{N\sin^2\theta}{16\pi} \big( -\{[Z^6,Z^8],\{Z^7,Z^8\}\}
  + \{[Z^7,Z^8],\{Z^6,Z^8\}\} \big) \notag
  \\
  &\quad\;\; = \frac{N\sin^2\theta}{16\pi}
  \begin{pmatrix}
    4c^{-}_M c^{+}_M + 4c^{-}_1 c^{+}_1 &&& \\
    & 4c^{-}_1 c^{+}_1 + 4c^{-}_2 c^{+}_2 && \\
    && \ddots && \\
    &&& 4c^{-}_{M-1} c^{+}_{M-1} + 4c^{-}_M c^{+}_M
  \end{pmatrix} , \notag
  \\
  &{Y^9}_8 \equiv \frac{i}{2\lambda} \{[X^9,X^6],B_{68}\} 
  + \frac{i}{2\lambda} \{[X^9,X^7],B_{78}\} \notag
  \\
  &\quad\;\; = \frac{N\sin^2\theta}{16\pi} \big( \{[Z^6,Z^9],\{Z^7,Z^9\}\}
  - \{[Z^7,Z^9],\{Z^6,Z^9\}\} \big) \notag
  \\
  &\quad\;\; = \frac{N\sin^2\theta}{16\pi}
  \begin{pmatrix}
    -4s^{-}_M s^{+}_M -4s^{-}_1 s^{+}_1 &&& \\
    & -4s^{-}_1 s^{+}_1 -4s^{-}_2 s^{+}_2 && \\
    && \ddots && \\
    &&& -4s^{-}_{M-1} s^{+}_{M-1} -4s^{-}_M s^{+}_M
  \end{pmatrix} , \notag
\end{alignat}
\begin{alignat}{3}
  {Y^6}_6 + {Y^7}_7 + {Y^8}_8 + {Y^9}_9
  &= - \frac{N}{\pi} \sin^2\theta \sin\tfrac{2\pi}{M} \mathbf{1}_{M} \notag
  \\
  &= -2\frac{N}{M} \sin^2\theta \mathbf{1}_{M} + \mathcal{O}
  \big((\tfrac{N}{M})^3 \big), 
  \\[0.2cm]
  ({Y^6}_6)^2 + ({Y^7}_7)^2 + \{{Y^6}_7,{Y^7}_6\}
  &= ({Y^8}_8)^2 + ({Y^8}_8)^2 + \{{Y^8}_9,{Y^9}_8\} \notag
  \\
  &= \bigg(\frac{N\sin^2\theta}{16\pi}\bigg)^2
  \big(32 \sin^2\tfrac{2\pi}{M} + 8 \sin^2\tfrac{4\pi}{M} \big)
  \mathbf{1}_{M} \notag
  \\
  &= \frac{N^2}{M^2} \sin^4\theta \mathbf{1}_{M} + \mathcal{O}
  \big((\tfrac{N}{M})^4 \big), \notag
  \\[0.1cm]
  \frac{f^2}{\lambda^2} \big( [X^6\!,\!X^8]^2 \!+\! [X^6\!,\!X^9]^2 
  \!+\! [X^7\!,\!X^8]^2 \!+\! [X^7\!,\!X^9]^2 \big)
  &= -\frac{N^2 \sin^2\theta \cos^2\theta}{2\pi^2}
  \big(1-\cos\tfrac{2\pi}{M}\big) \mathbf{1}_{M}, \notag
  \\
  &= -\frac{N^2}{M^2} \sin^2\theta \cos^2\theta \mathbf{1}_{M}
  + \mathcal{O} \big((\tfrac{N}{M})^4 \big). \notag
\end{alignat}
By using the above results, we obtain 
\begin{alignat}{3}
  \text{Tr}_i S &= {Y^6}_6 + {Y^7}_7 + {Y^8}_8 + {Y^9}_9 \notag
  \\
  &= -2\frac{N}{M} \sin^2\theta \mathbf{1}_{M} + \mathcal{O}
  \big((\tfrac{N}{M})^3 \big), \notag
  \\[0.2cm]
  -\tfrac{1}{2}\text{Tr}_i S^2 &= -\tfrac{1}{2} \Big( ({Y^6}_6)^2 \!+\! 
  ({Y^7}_7)^2 \!+\! \{{Y^6}_7,\!{Y^7}_6\} \!+\! ({Y^8}_8)^2 \!+\! 
  ({Y^8}_8)^2 \!+\! \{{Y^8}_9,\!{Y^9}_8\} \Big) 
  \\
  &\qquad -\frac{f^2}{\lambda^2} \big( [X^6\!,\!X^8]^2 \!+\! [X^6\!,\!X^9]^2 
  \!+\! [X^7\!,\!X^8]^2 \!+\! [X^7\!,\!X^9]^2 \big) \notag
  \\
  &= - \frac{N^2}{M^2} \sin^4\theta \mathbf{1}_{M}
  + \frac{N^2}{M^2} \sin^2\theta \cos^2\theta \mathbf{1}_{M}
  + \mathcal{O} \big((\tfrac{N}{M})^4 \big), \notag
  \\[0.2cm]
  \tfrac{1}{2} \big(\text{Tr}_i S \big)^2 &=
  2 \frac{N^2}{M^2} \sin^4\theta \mathbf{1}_{M}
  + \mathcal{O} \big((\tfrac{N}{M})^4\big). \notag
\end{alignat}

\end{document}